\documentclass[draftcls, onecolumn]{IEEEtran}

\usepackage{graphicx}
\usepackage[export]{adjustbox}
\usepackage{epstopdf}
\usepackage{cite}
\usepackage{latexsym}
\usepackage{amsmath, amssymb,amsfonts,amscd}
\usepackage{mathrsfs}
\usepackage{arydshln}
\usepackage{url}
\usepackage{enumerate}
\usepackage{subfig}
\usepackage{balance}
\usepackage[oldenum,olditem]{paralist}
\usepackage{caption}
\usepackage{color}
\usepackage{multirow}
\usepackage[utf8]{inputenc}
\usepackage[overload]{empheq}
\usepackage{algorithm2e}
\usepackage{algpseudocode}

\SetCommentSty{mycommfont}
\usepackage{units}
    
\newtheorem{theorem}{Theorem}[section]
\newtheorem{lemma}[theorem]{Lemma}

\newtheorem{definition}[theorem]{Definition}

\newcommand{\RR}{\mathbb{R}}

\newcommand{\mb}[1][]{\mathbf}
\newcommand{\m}[1]{\mb{m_{#1}}}
\newcommand{\x}{\mb{x}}

\newcommand{\bs}[1][]{\boldsymbol}

\newcommand{\eqn}{\begin{eqnarray}}
\newcommand{\feqn}{\end{eqnarray}}

\newcommand{\argmax}{\mathop{\rm arg~max}\limits}

\newcommand{\unaryminus}{{\tiny\scalebox{0.5}[1.0]{\( - \)}}}
\newcommand{\middleminus}{{\scalebox{0.7}[1.0]{\( - \)}}}
\newcommand{\middleequal}{{\scalebox{0.7}[1.0]{\( = \)}}}
\newcommand{\middledots}{{\scalebox{0.7}[1.0]{\( \dots \)}}}

\newcommand\restrict[1]{\raisebox{-.5ex}{$|$}_{#1}}

\usepackage{color}
\usepackage{wasysym}

\begin{document}

\title{A Geometrical-Statistical approach to outlier removal for {TDOA} measuments}

\author{M. Compagnoni, A. Pini, A. Canclini, P. Bestagini,~\IEEEmembership{Member,~IEEE}, F. Antonacci,~\IEEEmembership{Member,~IEEE},\\
S. Tubaro,~\IEEEmembership{Senior Member,~IEEE}, A. Sarti,~\IEEEmembership{Senior Member,~IEEE} 
\thanks{Marco Compagnoni and Alessia Pini are with the Dipartimento di Matematica, Politecnico di Milano, Piazza Leonardo da Vinci 32, 20133, Milan, Italy (e-mail: marco.compagnoni / alessia.pini@polimi.it).}
\thanks{Fabio Antonacci, Paolo Bestagini, Antonio Canclini, Stefano Tubaro and Augusto Sarti are with the Dipartimento di Elettronica Informazione e Bioingegneria, Politecnico di Milano, Piazza Leonardo da Vinci 32, 20133, Milan, Italy (e-mail: fabio.antonacci / paolo.bestagini / antonio.canclini / augusto.sarti / stefano.tubaro@polimi.it).}
}
\markboth{Journal of \LaTeX\ Class Files,~Vol.~11, No.~4, December~2012}%
{Shell \MakeLowercase{\textit{et al.}}: Bare Demo of IEEEtran.cls for Journals}

\maketitle

\begin{abstract}
The curse of outlier measurements in estimation problems is a well known issue in a variety of fields. Therefore, outlier removal procedures, which enables the identification of spurious measurements within a set, have been developed for many different scenarios and applications. In this paper, we propose a statistically motivated outlier removal algorithm for time differences of arrival (TDOAs), or equivalently range differences (RD), acquired at sensor arrays. The method exploits the TDOA-space formalism and works by only knowing  relative sensor positions. As the proposed method is completely independent from the application for which measurements are used, it can be reliably used to identify outliers within a set of TDOA/RD measurements in different fields (e.g. acoustic source localization, sensor synchronization, radar, remote sensing, etc.). The proposed outlier removal algorithm is validated by means of synthetic simulations and real experiments.
\end{abstract}

\begin{IEEEkeywords}
TDOA space, TDOA measurements, range differences, outlier removal.
\end{IEEEkeywords}

\IEEEpeerreviewmaketitle

\section{Introduction}\label{sec:intro}
Range Differences (RDs) are widely used in estimation problems based on sensor arrays, ranging from source localization to array calibration and synchronization \cite{Hasegawa2010,Koch95,Yimin2008,Pertila2013}. Given a pair of sensors, RD is defined as the difference between the two propagation distances from the emitting source to the sensors. RDs can be computed in several ways. For example, in wireless sensor networks RDs can be computed from Received Signal Strength Indications (RSSIs) measured at sensor pairs. In acoustic and radar applications, RDs are often obtained through peak-picking on the Generalized Cross Correlation (GCC) function \cite{Knapp1976} associated to the received signals. Notice that RDs are equivalent to Time Differences of Arrival (TDOAs), as they can be obtained by simply dividing the RD measurements by the propagation speed, which is normally assumed to be known and constant. This is why, with a slight abuse of terminology, hereafter we consider TDOAs, though our results can be readily extended to RDs. 

Given a distribution of $n+1$ sensors, the number of distinct possible TDOAs that can be measured is equal to $n(n+1)/2$ (complete set). Many applications, however, rely on a subset of $n$ TDOAs, all measured with respect to a reference sensor. This collection of $n$ TDOAs is typically referred to as the \textit{non-redundant set} or \textit{reduced set} as, in the absence of measurement noise, all the other TDOAs can be obtained through a linear combination of such measurements. The redundancy of the complete set, however, can be fruitfully exploited to achieve robustness of the considered application against additive measurement noise \cite{Cheung2008,Compagnoni2016a}.

Unfortunately, additive noise is only one of the problems that adversely affect solutions based on TDOAs. The presence of interfering sources or phenomena related to multipath propagation (i.e. reflection, diffusion and diffraction) tend to generate erroneous TDOAs (\emph{outliers}) and negatively affect the performance of TDOA-based applications if not properly accounted for.
Quite a few techniques have been developed for identifying and removing outliers in various areas of application (e.g. audio, radar, global-positioning systems, etc.). Such techniques can be broadly classified based on their working principle.
One class of solutions is based on an a-posteriori evaluation of the residuals of the used  cost function. More specifically, outliers are recognized as the measurements that contribute the most to the residual value of the cost function \cite{Picard2012,Tancredi2014,Jan1996,Picard2010}.
A second class of algorithms, specific for TDOA-based techniques, aims at improving the estimation of TDOAs by improving the GCC peak-picking strategy. In particular, heuristic rules related to the shape of the GCC in the proximity of peaks are applied in order to identify the TDOA that is most likely related to the desired source \cite{Canclini2013,Canclini2015,Bechler2005}.
A third class of algorithms is based on the observation that TDOAs must satisfy some geometrical and mathematical constraints based on signal propagation. A popular one is DATEMM \cite{Scheuing2006,Scheuing2008}, which is based on finding the sets of TDOAs that satisfy the raster and zero-sum conditions. The former compares the autocorrelation with the GCCs for discarding the GCC peaks generated by reflections. The latter aims at matching peaks in the GCC to each source that is present in the environment. 
It is worth noticing that the zero-sum condition exploits the fact that the sum of TDOAs over closed paths of three or more microphones (i.e., paths that begin and end at the same microphone) is bound to be zero.
In \cite{Velasco16} TDOA matrices (i.e. skew-symmetric matrices containing in the $ij$th position the TDOA between sensors $i$ and $j$) are found to satisfy some interesting algebraic properties, which are used for developing a set of denoising techniques in the presence of outliers. In particular, a two-step iterative algorithm is proposed, which requires the number of outliers to be small and upper-limited. 

In this work we propose a novel algorithm for removing outliers from a set of TDOA measurements. This method is not strictly related to the problem of source localization, as it does not depend on any localization cost function. Our algorithm is also independent of how TDOAs are measured, and it does not rely on a-priori information on the number of outliers within the measurement set. Our algorithm is based on the properties of the TDOA space, i.e. the space spanned by TDOA measurements. This space was introduced in \cite{Spencer2007} and further investigated in the literature. In particular, the closed-form analytical determination of the feasible set (i.e. the region in the TDOA space that corresponds to feasible source locations), and the mapping between TDOA and geometric spaces were subject of research in \cite{Compagnoni2013a} for the specific case of three sensors and source lying on the same plane. The analysis was accomplished for both complete and reduced TDOA sets. In the complete space, TDOAs are bound to lie on a plane. The equation of this plane corresponds to the zero-sum condition exploited in \cite{Scheuing2008}. Moreover, in the plane determined by the zero-sum condition, TDOAs corresponding to a real source must lie in a well defined closed region. In \cite{Compagnoni2016a} authors enriched this description with a statistical model and characterization.

In this contribution we exploit the concept of the feasible set derived from the TDOA space and we put the statistical model to work for the detection of outliers. More specifically, in a noiseless scenario, TDOAs are bound to lie within a well defined subset of the TDOA space. We show that, considering any single TDOA, this region is a finite interval in one dimension. Considering pairs of TDOAs that share a sensor, the feasible set is a convex region in two dimensions. Considering triplets belonging to a closed-loop of sensors the region is bound to lie on a convex subset of a given plane. In the presence of additive noise the considered group of TDOAs could fall outside the respective region. Nevertheless, it is possible to characterize the distance of the TDOAs from the corresponding feasible set exploiting the statistical noise model. In particular, assuming Gaussian additive noise, the distance from the feasible set must follow a Chi-Square distribution. An outlier removal algorithm is thus devised by testing the likelihood of the group of TDOAs (pairs and triplets) to be consistent with this Chi-Square distribution. The information coming from groups is then fused to identify and remove the outlier TDOAs, by means of an algorithm based on multiple testing and combined testing.
More specifically, multiple statistical tests are performed to detect the groups of TDOAs that potentially include outliers. The results of these tests are then combined to identify and remove the outlier TDOAs. We demonstrate that the devised algorithm yields a reliable outlier identification both in simulated and real scenarios. Moreover we also tested the proposed method in the context of acoustic source localization, whose accuracy is increased when outliers are removed from the set of TDOA measurements.

The rest of the manuscript is organized as follows. Section~\ref{sec:TS} introduces the TDOA space and defines the TDOA feasible regions when different numbers of sensors are considered. Section~\ref{sec:stat_models} explains how to formulate and test statistical hypotheses about the presence of outliers within TDOA sets exploiting the TDOA space formalism. Section~\ref{sec:outlier_removal} builds upon the previous one providing all the algorithmic details about the proposed TDOA outlier removal procedure. Section~\ref{sec:evaluation} describes a set of simulation results in order to validate the proposed approach. Section~\ref{sec:application} presents an example of application, which shows that the proposed algorithm helps improving TDOA-based source localization accuracy. Finally, Section~\ref{sec:conclusions} offers concluding remarks.
\section{The TDOA Space}\label{sec:TS}

In this section, we recall the definition and some useful properties of TDOA space, TDOA maps and feasible sets of the TDOAs, from \cite{Bestagini2013,Compagnoni2013a,Compagnoni2013b,Compagnoni2016a}. In particular, we give the full description of the feasible set when groups of two or three sensors are considered. We then describe some characteristics of the feasible set in the more general scenario of more than three sensors.

\subsection{TDOA map and feasible set}
Let us fix some ideas and notations that will be useful throughout the manuscript.
\begin{itemize}
\item
We identify the physical world with the 3D Euclidean space and, after choosing an orthogonal Cartesian coordinate system, with $\RR^3.$ The Euclidean scalar product of the vectors $\mb{v_1},\mb{v_2}$ and the norm of $\mb{v}$ are
$$
\langle\mb{v_1},\mb{v_2}\rangle_E=\mb{v_1}^T\mb{v_2}
\qquad\text{and}\qquad
\Vert\mb{v}\Vert_E=\sqrt{\langle\mb{v},\mb{v}\rangle_E}\ ,
$$
respectively.
\item
$\m{i}=(x_i,y_i,z_i)^T$ is the location of the $i$--th sensor. We take the indexes $i=0,\ldots,n$ and we assume that the $n+1$ sensors are in distinct positions.
\item
$\mb{d_{ji}}=\m{j}-\m{i}$ is the displacement vector from the sensor $\m{i}$ to the sensor $\m{j},$ for $i,j=0,\dots,n$.
\item
$\x$ is the position of the source.
\item
For notational simplicity, and with no loss of generality, we assume the propagation speed to be equal to 1, so that the noiseless TDOAs correspond to the range differences.
\end{itemize}

In this setting, we define the TDOA function for the pair of sensors $(\m{j},\m{i})$, $i\neq j$ as
\begin{equation}\label{eq:TDOAs1}
\begin{array}{cccc}
\tau_{ji}: & \RR^3 & \longrightarrow & \RR\\
& \x & \longmapsto & \tau_{ji}(\x)
\end{array},
\end{equation}
where
\begin{equation}\label{eq:TDOAs2}
\tau_{ji}(\x)=\Vert\x-\m{j}\Vert_E-\Vert\x-\m{i}\Vert_E.
\end{equation}
For every source position $\x$, the function $\tau_{ji}(\x)$ gives the value of the noiseless TDOA between the two selected sensors. We define the TDOA map as the function that collects all the $q=\frac{n(n+1)}{2}$ range differences \eqref{eq:TDOAs2} having $n\geq j>i\geq 0:$
\begin{equation}\label{eq:TDOAmap}\
\begin{array}{cccc}
\boldsymbol{\tau_n}: & \RR^3 & \longrightarrow & \RR^q\\
& \x & \longmapsto & (\tau_{10}(\x),\tau_{20}(\x),\ldots,\tau_{n\,n-1}(\x))^T
\end{array}\ .
\end{equation}
Therefore, for every source position $\x$, the map gives a $q$--dimensional vector that contains all the TDOAs with respect to $\x.$ In \cite{Compagnoni2013a}, $\boldsymbol{\tau_n}$ has been called the complete TDOA map, while the target set $\RR^q$ of $\boldsymbol{\tau_n}$ is referred to as the TDOA space.

The set of noiseless measurements generated by all the potential source positions coincides with the image $\text{Im}(\boldsymbol{\tau_n})$ of the TDOA map. We call it the feasible set $\Theta_n$. This means that any collection of noiseless TDOAs defines a point $\boldsymbol{{\tau}}= (\tau_{10},\ldots,\tau_{n\,n-1})^T\in\Theta_n$ and viceversa. However, a complete set of noisy TDOAs defines a point $\boldsymbol{\hat{\tau}}= (\hat{\tau}_{10},\ldots,\hat{\tau}_{n\,n-1})^T$ that is not necessarily in $\Theta_n.$ On the contrary, $\boldsymbol{\hat{\tau}}$ can be anywhere in the TDOA space $\RR^q.$ In particular, it is reasonable to assume that the presence of an outlier measurement $\hat{\tau}_{ji}$ can push the point $\boldsymbol{\hat{\tau}}$ really far away from $\Theta_n.$ The purpose of the manuscript is to delve deeper into this observation, in order to define a rigorous and statistically justified procedure able to correctly identify the erroneous coordinate $\hat{\tau}_{ji}$ of $\boldsymbol{\hat{\tau}}$ as an outlier. To this aim, it is necessary to gain a better understanding of the feasible sets $\Theta_n.$ This is the goal of the next subsections.

\subsection{The case $n=1$}\label{sec:n1}
\begin{figure}[htb]
\centering
\subfloat[][]{
  \resizebox{3.4cm}{!}{
  \includegraphics[valign=c]{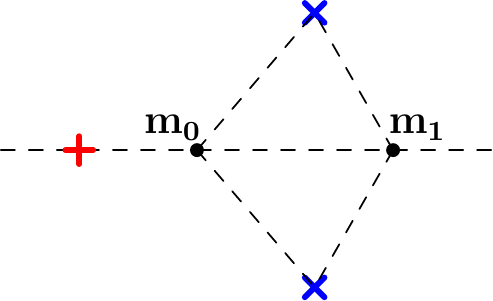}}
  \label{fig:tau1imagea}}\hspace{3mm}
\subfloat[][]{
  \resizebox{4.2cm}{!}{
  \includegraphics[valign=c]{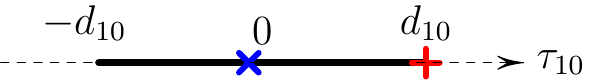}}\vspace{2mm}
  \label{fig:tau1imageb}}
  \caption{\label{fig:tau1image}
Crosses indicate the source positions on the left and the corresponding TDOAs on the right. The two symmetric blue sources generate the same blue TDOA. The image of $\bs{\tau_1}$ depicting the feasible set is the segment $\Theta_1=[-d_{10},d_{10}],$ contained in the linear space $V_1=\RR.$}
\end{figure}

In this situation the feasible set of TDOAs is very simple, as shown in Figure~\ref{fig:tau1image}. Two receivers $\m{0},\m{1}$ and a source $\x$ define a triangle, therefore $\tau_{10}(\x)$ is uniquely constrained by the corresponding triangular inequalities. It follows that
\begin{equation}\label{eq:imtau1}
\Theta_{1}=\{\tau_{10}\in\RR\,\vert\,-d_{10} \leq \tau_{10} \leq d_{10}\}\,.
\end{equation}
In Figure \ref{fig:tau1imagea} we show a configuration of two microphones $\m{0}=(0,0,0)^T,\m{1}=(1,0,0)^T$ and three different source positions $\x=(0.6,\pm 0.7,0)^T,(-0.6,0,0)^T$. In Figure \ref{fig:tau1imageb} the corresponding range difference measurements in the TDOA space are shown.

\subsection{The case $n=2$}\label{sec:n2}
Let us consider the function $\bs{\tau_2}\restrict{\RR^2}$ defined as the restriction of the TDOA map to $\RR^2.$ This map was thoroughly studied in \cite{Compagnoni2013a,Compagnoni2013b}, in the context of planar source localization.
Furthermore, in \cite{Compagnoni2016b} (Remark 10.4) is the proof that
$$
\Theta_2=\text{Im}(\bs{\tau_2})=\text{Im}(\bs{\tau_2}\restrict{\RR^2}).
$$
Here, we summarize the main properties of this set, that have been detailedly described in the above references.

$\Theta_2$ is a surface embedded into $\RR^3$, being the image of $\RR^2$ through $\bs{\tau_2}\restrict{\RR^2}$. Moreover, it is well known that the three TDOAs are not independent, since they satisfy the Zero-Sum Condition (ZSC) \cite{Scheuing2006}. Indeed, the linear relation $ \tau_{21}(\x ) = \tau_{20}(\x ) - \tau_{10}(\x ) $ holds for each $ \x \in \RR^3.$ Geometrically speaking, this means that three noiseless TDOAs are constrained on the plane
\begin{equation}
V_2=\{\bs{\tau}\in\RR^3\ |\ \tau_{10}-\tau_{20}+\tau_{21}=0\}\subset\RR^3
\end{equation}
and so $\Theta_2\subseteq V_2$.

Because of the above linear relation, in the literature it is customary to work with a reference sensor, for example $\m{0},$ and to consider only the two TDOAs $\tau_{10}(\x),\tau_{20}(\x).$ Mathematically speaking, let us define the reduced TDOA map
\begin{equation}
\begin{array}{cccc}
\bs{\tau_2^0}: & \RR^3 & \longrightarrow & \RR^2\\
& \x & \longmapsto & \quad
(\tau_{10}(\x),\tau_{20}(\x))^T
\end{array}.
\end{equation}
The image $\Theta_2^0=\text{Im}(\bs{\tau_2^0})$ is strictly related to the feasible set $\Theta_2.$ Indeed, let us take the forgetting map
\begin{equation}\label{eq:forgettingmap}\
\begin{array}{cccc}
p_3: & \RR^3 & \longrightarrow & \RR^2\\
& (\tau_{10},\tau_{20},\tau_{21})^T & \longmapsto & (\tau_{10},\tau_{20})^T
\end{array}\ ,
\end{equation}
i.e. the projection that takes care of forgetting the third coordinate. Then, we have $\bs{\tau_2^0}=p_3\circ\bs{\tau_2},$ where $\circ$ is the composition operator. Moreover, $p_3$ is a natural bijection between the feasible set $\Theta_2$ and $\Theta_2^0,$ as one can see in Figure~\ref{fig:taucomimage}. 
Hence, we can investigate the properties of the noiseless TDOA model by studying the simpler map $\bs{\tau_2^0}$. From now on, we separately consider the cases of sensors in general and aligned configurations.

\subsubsection{General configurations}\label{sec:genconf}
\begin{figure}[htb]
\centering
\subfloat[][]{
  \resizebox{4.3cm}{!}{
  \includegraphics[valign=c]{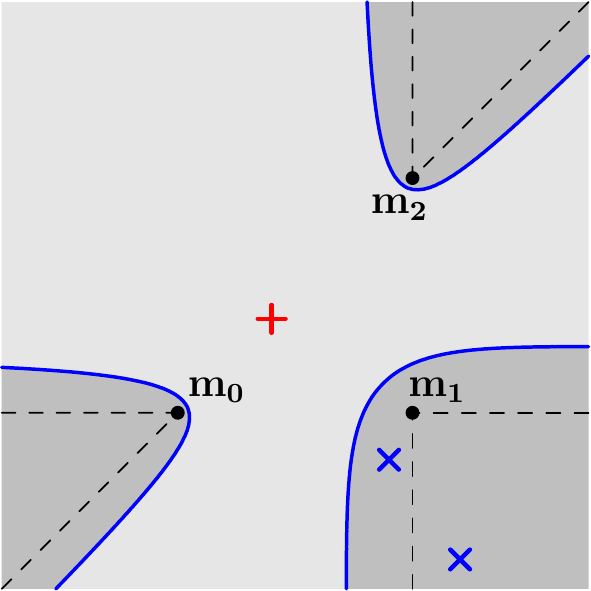}}
  \label{fig:tauimagea}}\\
\hspace{0mm}\subfloat[][]{
  \resizebox{6.7cm}{!}{
  \includegraphics[valign=c]{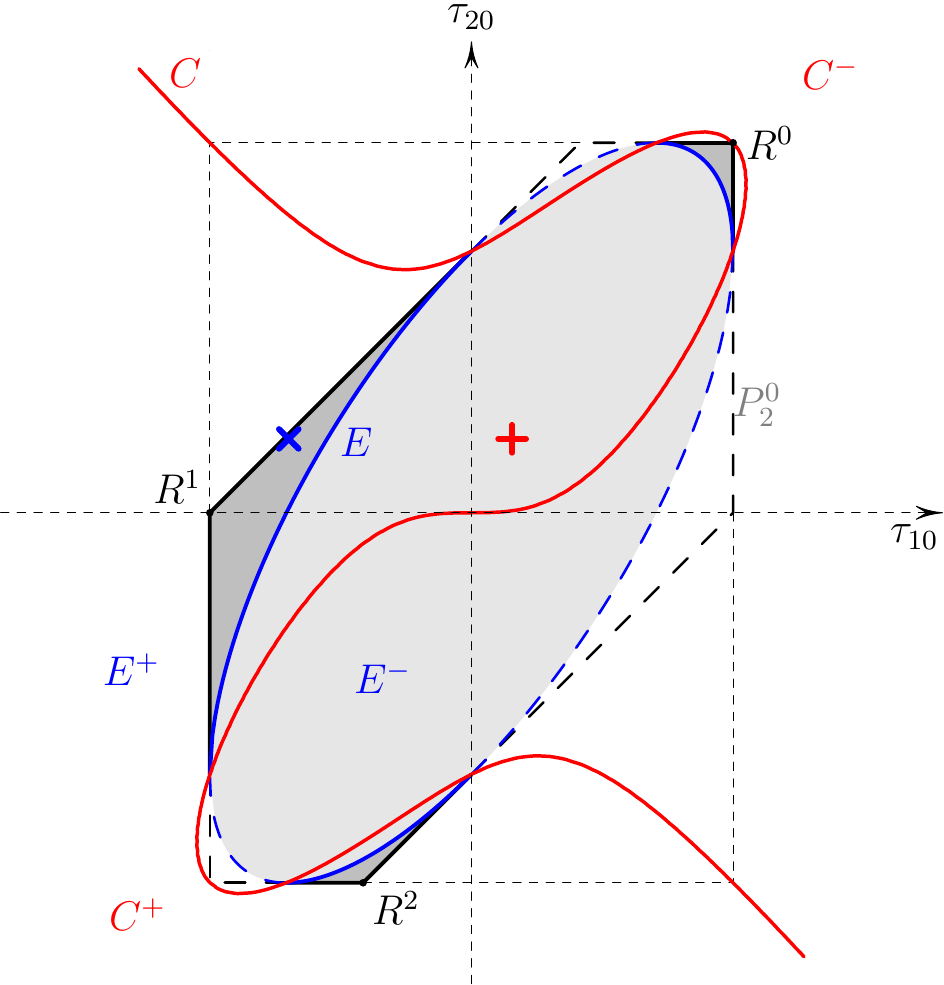}}\vspace{2mm}
  \label{fig:tauimageb}}
  \caption{\label{fig:tauimage}
The crosses indicate the source positions in the physical space and the corresponding TDOAs in $\Theta_2^0$. The two blue sources generate the same blue TDOA. The image of $\bs{\tau_2^0}$ is the gray subset of the hexagon $P_2^0$ with continuous and dashed sides. Let us observe that the medium gray region is entirely contained in $C^+\cap P_2^0.$ The continuous part of the boundary of the hexagon and the blue ellipse $E,$ together with the vertices $R^i,$ are in the image, while the dashed boundaries do not belong to $\Theta_2^0$.}
\end{figure}

First of all, we observe that $\Theta_2^0$ is contained into the hexagon $P_2^0$ defined by the following triangle inequalities:
\begin{equation}\label{eq:politopo}
\left\{ \begin{array}{l}
-d_{10} \leq \tau_{10} \leq d_{10} \\
-d_{20} \leq \tau_{20} \leq d_{20} \\
-d_{21} \leq \tau_{20} - \tau_{10} \leq d_{21}
\end{array} \right. .
\end{equation}
In particular, the vertices of the hexagon $R^0=(d_{10},d_{20})^T,\ R^1=(-d_{10},d_{21}-d_{10})^T,\ R^2=(d_{21}-d_{20},-d_{20})^T$ of $P_2^0$ correspond to the pairs of TDOAs associated to a source at $\m{0},\m{1},\m{2},$ respectively (see Figure \ref{fig:tauimage}).

Without loss of generality, we assume that the receivers lie on the horizontal plane and the displacement vectors $\mb{d_{10}},\mb{d_{20}}$ are counterclockwise oriented. Let us consider $\bs{\tau^0}=(\tau_{10},\tau_{20})^T\in\RR^2.$ Following the analysis of Section 6 in \cite{Compagnoni2013a}, we give the following definition.
\begin{itemize}
\item
$
W=\det\left(
\begin{array}{ccc}
\mb{d_{10}} & \mb{d_{20}} & \mb{e_3}
\end{array}\right)
$
and the rotation matrix
$$
\mb{H}=\left(
\begin{array}{ccc}
0 & -1 & 0\\
1 & 0 & 0\\
0 & 0 & 1
\end{array}\right).
$$
\item
The vectors
\begin{equation}
\mb{v}(\bs{\tau^0})=\mb{H}\,(\tau_{20}\, \mb{d_{10}} - \tau_{10}\, \mb{d_{20}})\,,
\end{equation}\\[-6mm]
\begin{equation}
\mb{l_0}(\bs{\tau^0})=\displaystyle
\mb{H}\,\frac{(d_{20}^2-\tau_{20}^2)\,\mb{d_{10}} -
(d_{10}^2-\tau_{10}^2)\,\mb{d_{20}}}
{2\, W}.
\end{equation}
\item
The polynomials
\begin{equation}
\begin{array}{l}
a(\bs{\tau^0}) = \Vert\mb{v}(\bs{\tau^0})\Vert^2_E -W^2,\\[1mm]
b(\bs{\tau^0}) = \langle\mb{v}(\bs{\tau^0}),\mb{l_0}(\bs{\tau^0})\rangle_E.
\end{array}
\end{equation}
\end{itemize}
What we obtain are the following facts:
\begin{itemize}
\item
$a(\bs{\tau^0})=0$ is the equation of the unique ellipse $E$ tangent to every facet of $P_2^0.$ We name $E^-$ and $E^+$ the interior and the exterior regions of $E$ where $a(\bs{\tau^0})<0$ and $a(\bs{\tau^0})>0,$ respectively;
\item
$b(\bs{\tau^0})=0$ is the equation of a cubic curve $C.$ We name $C^-$ and $C^+$ the regions where $b(\bs{\tau^0})<0$ and $b(\bs{\tau^0})>0,$ respectively.
\end{itemize}
In \cite{Compagnoni2013a} is a proof of the fact that the image of $\bs{\tau_2^0}$ is given by the set
\begin{equation}
\Theta_2^0 = E^- \cup (C^+\cap P_2^0) \cup R^0.
\end{equation}

\begin{figure}[htb]
	\begin{center}
		\resizebox{6.5cm}{!}{
			\includegraphics[trim={0 2.5cm 0 3.5cm},clip]{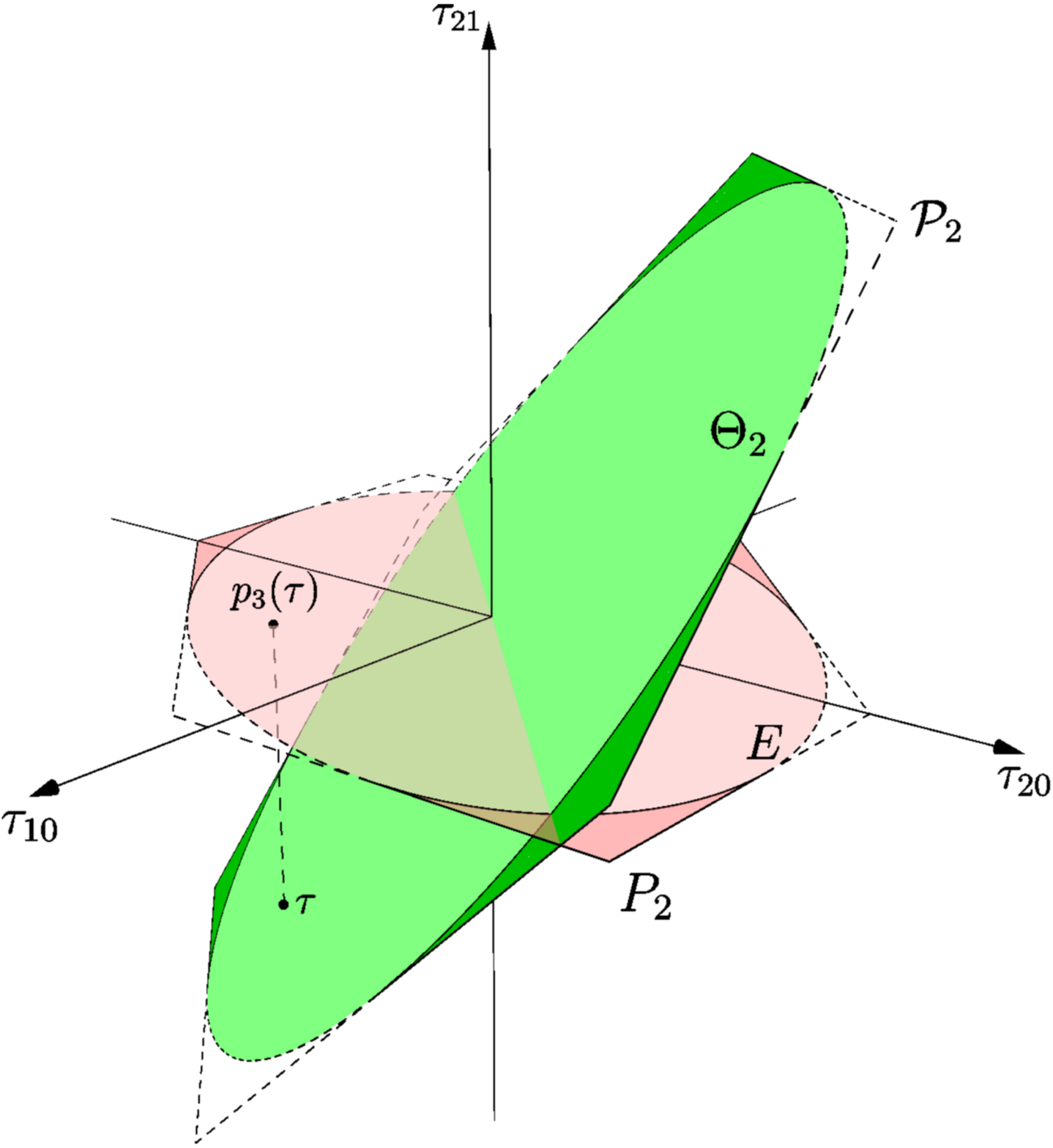}}
		\caption{\label{fig:taucomimage}The image of $\bs{\tau_2}$ is the green subset of the hexagon $p_3^{-1}(P_2^0)\subset V_2,$ while the image of $ \bs{\tau_2^0}$ is the red subset of $P_2^0.$ There is a 1--to--1 correspondence between $\Theta_2$ and $\Theta_2^0$ via the map $p_3$.
		}
	\end{center}
\end{figure}

In Figure \ref{fig:tauimagea} we show the plane containing the sensors at $\m{0}=(0,0,0)^T,\m{1}=(1,0,0)^T,\m{2}=(1,1,0)^T$ and three source positions at $\x=(0.4,0.4,0)^T,(0.9,-0.2,0)^T$ and approximately $(1.2,-0.625,0)^T$. In Figure \ref{fig:tauimageb} it is depicted the corresponding situation in $\Theta_2^0.$ Roughly speaking, the set $\Theta_2^0$ is given by the interior of the ellipse and the three medium gray regions. These three subsets of $P_2^0$ are on the same side of the curve $C,$ in the region $C^+.$ The TDOA map is a bijection between the light gray regions in the physical and TDOA spaces, while it is $2$--to--$1$ in the medium gray regions (the interested reader can find the complete analysis of these properties in \cite{Compagnoni2013a,Compagnoni2013b}). Finally, Figure \ref{fig:taucomimage} shows the relation between $\Theta_2$ and $\Theta_2^0$ given by the forgetting map.

\subsubsection{Aligned configurations}\label{sec:alconf}
Let us assume that $\m{0},\m{1},\m{2}$ are aligned, then two of the inequalities \eqref{eq:politopo} are redundant and the polyhedron $P_2^0$ is the parallelogram with vertices $R^0,R^1,R^2.$
The image of $\bs{\tau_2^0}$ is the triangle $T$ with the same vertices minus the diagonal of the parallelogram.

\begin{figure}[htb]
\centering
\subfloat[][]{
  \resizebox{3.5cm}{!}{
  \includegraphics[valign=c]{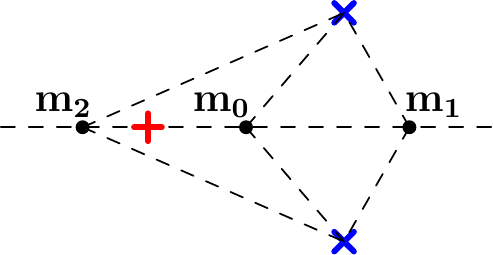}}
  \label{fig:tauimagealigneda}}\hspace{3mm}
\subfloat[][]{
  \resizebox{4.2cm}{!}{
  \includegraphics[valign=c]{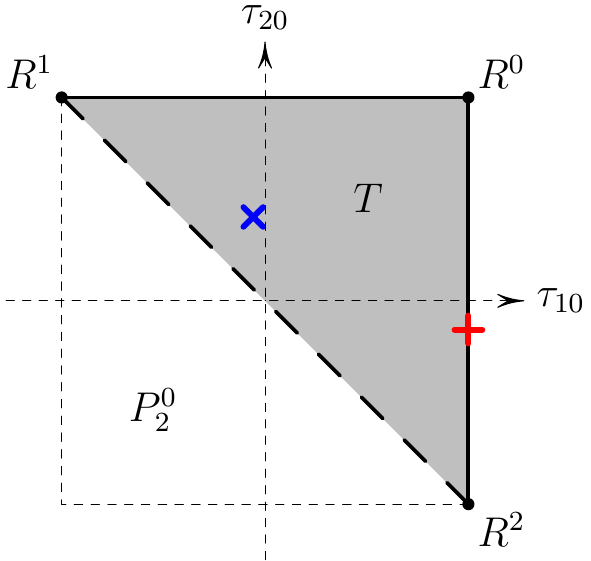}}\vspace{2mm}
  \label{fig:tauimagealignedb}}
  \caption{\label{fig:tauimagealigned}
The crosses indicate the source positions on the left and the corresponding TDOAs on the right, where there is the image of $\bs{\tau_2^0}.$ The polyhedron $P_2^0$ is a rectangle.
The dashed side of $T$ is not in $\Theta_2^0$.}
\end{figure}

In Figure \ref{fig:tauimagealigneda} we show a configuration of three aligned sensors $\m{0}=(0,0,0)^T,\ \m{1}=(1,0,0)^T\ \text{and}\ \m{2}=(-1,0,0)^T$ and three different source positions $\x=(0.6,\pm0.7,0)^T,(-0.6,0,0)^T.$ In Figure \ref{fig:tauimagealignedb} there is the set $\Theta_2^0$ and the TDOAs associated to the source positions on the left.

For future reference, we give the inequalities defining the topological closure of $T.$ Let us define $c^{\pm\pm\pm}=\pm d_{10}\pm d_{20}\pm d_{21},$ then:
\begin{equation}\label{eq:parallelogramma}
\left\{\begin{array}{l}
c^{--+}\,\tau_{10}+2\,d_{10}\,\tau_{20}\leq d_{10}\,c^{-++}\\
2\,d_{20}\,\tau_{10}+c^{--+}\,\tau_{20}\leq d_{20}\,c^{+-+}\\
c^{+--}\,\tau_{10}+c^{-+-}\,\tau_{20}\leq d_{21}\,c^{++-}
\end{array}\right. .
\end{equation}

\subsection{The general case $(n>2)$}\label{sec:generalTDOAspace}
We do not have the full description of $\boldsymbol{\tau_n}$ for $n>2$ and its study goes beyond the scope of this manuscript. Instead, in this section we focus on a simpler problem, which is the analysis of the linear relations that exist between the $q$ TDOAs of the complete set. We begin our analysis by introducing the space of the linear relations among the TDOAs.
\begin{definition}\label{def:linearrelations}\emph{
Let us consider $n+1$ sensors at $\m{0},\ldots,\m{n}$ in $\RR^3,$ where $\mathrm{n}\geq 2$.
\begin{enumerate}[i)]
\item
A linear relation $l(a_{10},\dots,a_{n\,n-1})$ of the TDOAs is a linear combination of the TDOA functions that is identically equal to zero. This means that
\begin{equation}\label{eq:linearrelations}
l(a_{10},\dots,a_{n\,n-1})=\sum_{i=0}^{n-1}\sum_{j=i+1}^n a_{ji}\tau_{ji}(\x),
\end{equation}
where $a_{ji}\in\RR$ with $0\leq i<j\leq n,$ and $l(a_{10},\dots,a_{n\,n-1})=0$ for every $\x\in\RR^3.$
\item
The {\em length} of $l(a_{10},\dots,a_{n\,n-1})$ is the number of non-zero coefficients $a_{ji}.$
\item
We name $\mathcal{L}_n$ the set of all linear relations.
\end{enumerate}
}\end{definition}

ZSCs are an example of linear relations. In particular, we say that a ZSC is minimal if it involves only three sensors, or equivalently, if its length is three:
\begin{equation}\label{eq:linearrelationsZSC}
\tau_{ji}(\x)-\tau_{ki}(\x)+\tau_{kj}(\x)=0,\quad 0\leq i<j<k\leq n.
\end{equation}

A geometric interpretation of the linear relations between the TDOAs can be found in \cite{Compagnoni2016a}. Indeed, they define an $n$--dimensional linear subspace $V_n$ of the TDOA space $\RR^q$, and $\Theta_n$ is a complicated three-dimensional semialgebraic variety (i.e. a set defined by algebraic equations and inequalities \cite{Basu2006}) embedded in $V_n.$

As we will see in the next Sections, our algorithm for the removal of the outliers among TDOA measurements is based on the linear relations that are valid in minimal sets of TDOAs.
With this in mind, the aim of the section is to show that the minimal ZSCs play a key role among the possible linear relations between TDOAs:
\begin{enumerate}[a)]
\item\label{enum:a}
the minimal ZSCs are exactly the relations that involve the minimum number of TDOAs;
\item\label{enum:b}
any other relation can be decomposed into a linear combination of minimal ZSCs.
\end{enumerate}

To this purpose, we investigate in depth the properties of $\mathcal{L}_n.$ There is a natural linear structure over this set. Indeed, it is straightforward to verify that $\mathcal{L}_n$ is closed under the operations
\begin{itemize}
\item
$l(a_{10},\dots,a_{n\,n-1})+l(b_{10},\dots,b_{n\,n-1})=l(a_{10}+b_{10},\dots,a_{n\,n-1}+b_{n\,n-1});$
\item
$t\cdot l(a_{10},\dots,a_{n\,n-1})=l(ta_{10},\dots,ta_{n\,n-1})\ $ for $\ t\in\RR$
\end{itemize}
and that they satisfy all the requirements for vector spaces. Therefore, $\mathcal{L}_n$ is a real vector space.

Then, we can state the main theorem of this subsection, that characterizes the space $\mathcal{L}_n$ and the role of ZSCs.
\begin{theorem}\label{th:Vn}\emph{
\begin{enumerate}[i)]
\item
the minimal length of a non-trivial linear relation $l(a_{10},\dots,a_{n-1\, n})\in \mathcal{L}_n$ is three;
\item
the set of length-three relations coincides with the set of minimal ZSCs, up to multiplicative constants;
\item
a basis of $\mathcal{L}_n$ is given by the minimal ZSCs involving a reference sensor, for example $\m{0}$
\begin{equation}\label{eq:linearrelationsZSCm0}
\tau_{i0}(\x)-\tau_{j0}(\x)+\tau_{ji}(\x)=0,\quad 0<i<j\leq n.
\end{equation}
\end{enumerate}
}\end{theorem}
Notice that properties \eqref{enum:a} and \eqref{enum:b} of the minimal ZSCs are a direct consequence of Theorem \ref{th:Vn}. In the next section we will analyze their consequences for our outlier removal procedure.

We conclude our investigation by proving the above theorem. We need the following lemma.
\begin{lemma}\label{lm:toacomb}\emph{
An equation in $\x\in\RR^3$ of the form
\begin{equation}\label{eq:toacomb}
\displaystyle \sum_{i=0}^{n}\,\omega_i\,\Vert\x-\m{i}\Vert=0,\qquad \omega_i\in\RR,
\end{equation}
is an identity, i.e. it is satisfied for every $\x,$ if and only if $\omega_0=\dots=\omega_n=0.$ In general, equation \eqref{eq:toacomb}
defines a set in $\RR^3$ that is contained in an algebraic surface.
}\end{lemma}
\noindent\emph{Proof of Lemma \ref{lm:toacomb}:}
the lemma follows by adapting the proof of Theorem 4.1. in \cite{Nie2008}.
\hfill$\square$\vspace{1mm}

\noindent\emph{Proof of Theorem \ref{th:Vn}:}
Let us define $a_{ii}=0$ and $a_{ij}=-a_{ji}.$ Thus, we can rewrite the linear form in \eqref{eq:linearrelations} as
\begin{equation*}
\begin{array}{l}
\!l(a_{10},\dots,a_{n-1\, n})=
\displaystyle \sum_{i=0}^{n-1}\sum_{j=i+1}^n a_{ji}\,(\Vert\x-\m{j}\Vert-\Vert\x-\m{i}\Vert)\\
\phantom{l(a_{10},\dots,a_{n-1\, n})}=
\displaystyle \sum_{j=0}^{n}\left(\sum_{i=0}^{n}\,a_{ji}\right)\Vert\x-\m{j}\Vert.\\
\end{array}
\end{equation*}
By Lemma \ref{lm:toacomb}, the linear relation $l(a_{10},\dots,a_{n-1\, n})=0$ is true if and only if
\begin{equation}\label{eq:linearrelations2}
\sum_{i=0}^{n}\,a_{ji}=0\quad \text{for each}\quad j=0,\dots,n.
\end{equation}
Let us define the skew-symmetric matrix
$$
\mb{A}\;\middleequal
\begin{pmatrix}
a_{00} & a_{01} & \middledots & a_{0n}\\
a_{10} & a_{11} & \middledots & a_{1n}\\
\middledots & \middledots & \middledots & \middledots\\
a_{n0} & a_{n1} & \middledots & a_{nn}
\end{pmatrix}\middleequal
\begin{pmatrix}
0 & \middleminus a_{10} & \middledots & \middleminus a_{n0}\\
a_{10} & 0 & \middledots & \middleminus a_{n1}\\
\middledots & \middledots & \middledots & \middledots\\
a_{n0} & a_{n1} & \middledots & 0
\end{pmatrix}.
$$
Recalling \eqref{eq:linearrelations2}, the coefficients of $\mb{A}$ define a linear relation if and only if the vector $\mb{1}=(1\ \dots\ 1)^T$ lies on $\ker(\mb{A})$. We now use the structure of the matrix $\mb{A}$ to prove the theorem.

It should be quite clear that the length of any non-trivial relation is at least two. With no loss of generality, let us assume that $a_{10}\neq 0.$ From the product of the first row of $\mb{A}$ and $\mb{1},$ the second non-zero coefficient of $\mb{A}$ must be $a_{j0}=-a_{10}$ for some $j=2,\dots,n.$ We can assume $a_{20}=-a_{10}$. This, however, implies that the product of the second and third row of $\mb{A}$ and $\mb{1}$ are $a_{10}\neq 0$ and $a_{20}\neq 0$, respectively, hence $\mb{1}\notin\ker(\mb{A})$. This is why we have to consider length-three relations. In this case, it is easy to check that there is only one way to set both the above products to zero, which is by setting $a_{21}=a_{10}=-a_{20}$. We therefore obtain a ZSC (up to a scaling factor), which proves the first two claims of the theorem.

In order to prove the third claim, we begin with noticing that \eqref{eq:linearrelations2} is a linear system of $n+1$ equations in $q$ variables. The first $n$ equations are clearly independent, while the last one is a linear combination of the others. This follows from the identity $\mb{1}^T\mb{A}\mb{1}=0$, which is, in turn, a consequence of the skew symmetry of $\mb{A}$. The dimension of $\mathcal{L}_n$ is, therefore, $q-n$. As the ZSCs (16) form an independent set of $q-n$ linear relations,
they define a basis of $\mathcal{L}_n.$
\hfill$\square$\vspace{1mm}

We finally remark that the linear relations in $\mathcal{L}_n$ do not depend on the position of the receivers. This is in contrast, for example, with both the inequalities \eqref{eq:imtau1} and (\ref{eq:politopo},\ref{eq:parallelogramma}), which constrain $\Theta_1$ and $\Theta_2^0$, respectively. In particular, this means that they are not sensitive to the accuracy in the knowledge about the location of sensors.
\section{Statistical tests on $G$-tuples of TDOAs}\label{sec:stat_models}
In the presence of measurement errors on the data, we must resort to statistical modeling. Using the same notations of Section \ref{sec:TS}, let us consider $n+1$ sensors. By assuming additive Gaussian noise, the TDOAs associated to a source in $\x$ are described by the multivariate parametric model \cite{Benesty2004,Cheung2008}
\begin{equation}\label{eq:TDOAstatmod}
\bs{\hat{\tau}_n}(\x)=\bs{\tau_n}(\x)+\bs{\varepsilon},
\end{equation}
where $\bs{\varepsilon}\sim N(\mb{0},\bs\Sigma)$. Additive error could be, as an example, the error introduced by acquiring at a finite sampling frequency.
An outlier is assumed as a measurement that is not compatible with \eqref{eq:TDOAstatmod}. In order to identify the outlier, we work in a two-step fashion. In this section we focus on the problem of testing whether a $G$-tuple of TDOAs, $G=1,2,3$, potentially contains outliers. This information will then be used for outlier detection and removal as explained in Section~\ref{sec:outlier_removal}.

\subsection{Statistical noise model}
Under the assumption \eqref{eq:TDOAstatmod} and without the presence of outliers, the probability density function (p.d.f.) of the TDOA set is
\begin{equation}\label{eq:TDOAprob}
p(\bs{\hat{\tau}};\bs{\tau_n}(\mb{x}),\bs\Sigma)=
\frac{e^{-\frac{1}{2}(\bs{\hat{\tau}}-\bs{\tau_n}(\mb{x}))^T \bs\Sigma^{-1}(\bs{\hat{\tau}}-\bs{\tau_n}(\mb{x}))}}{\sqrt{(2\pi)^q|\bs\Sigma|}}\,.
\end{equation}
From a geometric standpoint (see, for example, \cite{Amari2000}), the Fisher matrix $\bs\Sigma^{-1}$ defines a Euclidean structure on $\RR^q$. In general, given a symmetric and positive-definite matrix $\mb{G}$ of order $q$, it allows us to define the scalar product
\begin{equation}\label{eq:Fisherscalarprod}
\langle\mb{v_1},\mb{v_2}\rangle_{\mb{G}} = \mb{v_1}^T \mb{G} \mb{v_2},\qquad \mb{v_1},\mb{v_2}\in\RR^q
\end{equation}
on the vector space $\RR^q$ and the corresponding norm
\begin{equation}\label{eq:Mahalanobis}
\Vert\mb{v}\Vert_{\mb{G}} =\sqrt{\mb{v}^T \mb{G} \mb{v}}\,,\qquad \mb{v}\in\RR^q \; .
\end{equation}
If $\mb{G}=\bs\Sigma^{-1}$, \eqref{eq:Fisherscalarprod} and \eqref{eq:Mahalanobis} are known in the statistical literature as the Mahalanobis product and norm, respectively.
Within this setting, the p.d.f. \eqref{eq:TDOAprob} can be rewritten as
\begin{equation}\label{eq:TDOAprob2}
p(\bs{\hat{\tau}};\bs{\tau_n}(\mb{x}),\bs\Sigma)=
\frac{e^{-\frac{1}{2}\Vert\bs{\hat{\tau}}-\bs{\tau_n}(\mb{x})
		\Vert_{\bs\Sigma^{\unaryminus 1}}^2}}{\sqrt{(2\pi)^q|\bs\Sigma|}}\,,
\end{equation}
which only depends on the square of the Mahalanobis distance between $\bs{\hat{\tau}}$ and $\bs{\tau_n}(\mb{x}).$

As already noted in Section \ref{sec:TS}, in the presence of noisy measurements the point $\bs{\hat{\tau}}$ can be outside the feasible set $\Theta_n.$ In particular, a set of measurements containing an outlier may define a point very distant from $\Theta_n.$ Under the assumption \eqref{eq:TDOAstatmod}, the squared Mahalanobis distance between $\bs{\hat{\tau}}$ and $\bs{\tau_n}(\mb{x})$ has a Chi--square distribution with $q$ degrees of freedom \cite{anderson1962}:
$$
d(\bs{\hat{\tau}},\bs{\tau_n}(\x);\bs\Sigma)^2=
\Vert\bs{\hat{\tau}}-\bs{\tau_n}(\x)\Vert_{\bs\Sigma^{\unaryminus 1}}^2
\sim \chi^2_q\,.
$$
However, as we generally do not know the position $\x$ of the source, we have to focus on the distance of $\bs{\hat{\tau}}$ from $\Theta_n$. 
In the following, we discuss in details the cases involving sets composed by one, two and three TDOAs.

\subsection{$G=1$: Test on single TDOAs}\label{subsec:k_1}
Let us consider the null hypothesis $H_{0,ji}$ that $\hat{\tau}_{ji}$ is not an outlier and the corresponding alternative hypothesis $H_{1,ji}=H_{0,ji}^C.$ Our goal is to build a test for assessing which one of the hypotheses is correct. In particular - following common statistical practice - we control the probability of type-I errors, i.e. the probability of wrongly rejecting $H_{0,ji}$ in favor of $H_{1,ji}$:
\begin{equation}\label{eq:exact.test.1D}
P[\text{reject } H_{0,ji} | H_{0,ji} \text{ true}] \leq \alpha,
\end{equation}
$\alpha$ being a pre-defined probability value (typically $5\%$). With this goal in mind, we define a test statistic and compute a $p$-value $\lambda_{ij}$, which, under the null hypothesis $H_{0,ji}$, measures the probability that the statistic be equal to or larger than its observed value. If the $p$-value of the test is larger than $\alpha$, the null hypothesis is accepted, otherwise it is rejected. 

Let us call $\Theta_{1,ji}$ the feasible set of the single TDOA $\hat{\tau}_{ji}.$ As seen in Section \ref{sec:TS}, we have $\Theta_{1,ji}=[-d_{ji},d_{ji}]$. A natural choice for the test statistic is the squared Mahalanobis distance $d(\hat\tau_{ji},\Theta_{1,ji};\sigma_{ji})^2.$
We have:
\begin{equation}\label{eq:1Ddist}
d(\hat\tau_{ji},\Theta_{1,ji};\sigma_{ji})=
\left\{\begin{array}{ll}
0 & \text{if}\quad\hat\tau_{ji}\in\Theta_{1,ji}\,,\\
\displaystyle\frac{|\hat\tau_{ji}|-d_{ji}}{\sigma_{ji}} & \text{otherwise.}
\end{array}\right.
\end{equation}
To compute the $p$-value of the test we need to find the distribution of the statistic under the null hypothesis. Unfortunately, this depends on the true parameter $\tau_{ji},$ that is unspecified  and is only known to lie in $\Theta_{1,ji}$. On the other hand, $P[\text{reject } H_{0,ji} | H_{0,ji} \text{ true}]$ is maximized when $\tau_{ji}$ lies on the boundary of the feasible set, e.g. if $\tau_{ji} = d_{ji}$. Hence, in order to compute $\lambda_{ji}$ we put us in this extreme situation, so that $P[\text{reject } H_{0,ji} | H_{0,ji} \text{ true}] \leq \alpha$ regardless of the unknown value $\tau_{ji}$.

Based on the analysis in \cite{Chernoff1954,self1987asymptotic}, we have the following consequential facts:
\begin{enumerate}[i.]
\item
if the noise $\sigma_{ij}$ is negligible with respect to $d_{ji},$ we can approximate $\Theta_{1,ji}$ with the half--line $(-\infty,d_{ji}];$
\item
the squared Mahalanobis distance $d(\hat\tau_{ji},\Theta_{1,ji};\sigma_{ji})^2$ has the same probability to be equal to $0$ or to follow a Chi-square distribution with one degree of freedom. So, the test statistic distribution is a mixture between a Dirac delta on zero (that we denote with $\chi^2_0$), and a $\chi^2_1$:
$$
d(\hat\tau_{ji},\Theta_{1,ji};\sigma_{ji})^2 \sim \frac{1}{2} \chi^2_0 + \frac{1}{2} \chi^2_1;
$$
\item
the $p$-value of the test can be computed as
$$
\lambda_{ji} = \frac{1}{2}\left(1 - F_{\chi^2_1}( d(\hat\tau_{ji},\Theta_{1,ji};\sigma_{ji})^2) \right),
$$
where $F_{\chi^2_1}$ is a Chi-square cumulative distribution function (c.d.f.) with one degree of freedom.
\end{enumerate}

\begin{figure}[htb]
\begin{center}
\resizebox{5cm}{!}{
  \includegraphics{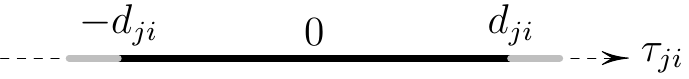}}\vspace{2mm}
  \caption{\label{fig:tau1imageout}
In the preprocessing of the TDOAs, we eliminate every $\hat\tau_{ji}$ that is not sufficiently close to the feasible set $\Theta_{1,ji}=[-d_{ji},d_{ji}]$. The acceptance region $\mathcal{I}_{ji}$ corresponds to $\Theta_{1,ji}$ plus the grey segments. A TDOA lying on one of them is not considered an outlier, even though it would not be feasible in a noiseless scenario.}
\end{center}
\end{figure}

The measurements $\hat\tau_{ji}$ whose  $p$-values $\lambda_{ji}$ are lower than $\alpha$ are marked as outliers, and removed from the set of measurements. By setting $\alpha=5\%$, we are guaranteed that the probability of wrongly removing an inlier is $5\%$ in the worst case. At the end, we are able to define the so-called acceptance region for $\hat\tau_{ji}$ as
\begin{equation}
	\label{eq:outlier_interval}
	\mathcal{I}_{ji} = [-d_{ji}-\gamma_{2\alpha,ji}, \; d_{ji}+\gamma_{2\alpha,ji}],
\end{equation}
where $\gamma_{2\alpha,ji} = \sigma_{ji} \sqrt{F_{\chi^2_1}^{-1}(1-2\alpha)}$. The TDOA is considered an outlier if $\hat\tau_{ji}\notin \mathcal{I}_{ji}$ (see Figure \ref{fig:tau1imageout}).

The single TDOA test is a refinement of the usual practice of removing every $\hat{\tau}_{ji}$ that does not satisfy the corresponding triangular inequality \cite{Scheuing2006}. As a matter of fact, we can consider it as a form of preprocessing of the measured data and in the rest of the manuscript we will assume to always perform it.

\subsection{$G=2$: Test on pairs of TDOAs }

Let us consider a pair of TDOAs $\hat{\tau}_{ji},\hat{\tau}_{ki}$ estimated from the signals acquired by a shared sensor of index $i\in\{0,\dots,n\}$ and two other sensors of indices $j$, $k$, where $0\leq j<k\leq n$ and $k,j\neq i$. The TDOAs define the point $\bs{\hat{\tau}_{kj}^i}=(\hat{\tau}_{ji},\hat{\tau}_{ki})^T\in\RR^2,$ whose corresponding feasible set $\Theta_{2,kj}^i$ has been described in Sections \ref{sec:genconf} and \ref{sec:alconf}, for the case $i=0,\ j=1,\ k=2$.

For investigating the presence of at least one outlier in $\bs{\hat{\tau}_{kj}^i}$, we proceed as in the previous case. For each considered microphone pair, we need to construct a statistical test on the hypotheses
\begin{equation}\label{eq:test_couple}
	H_{0,kj}^i : \bs{\hat\tau_{kj}^i} \text{ does not contain outliers;}\   
	H_{1,kj}^i = H_{0,kj}^{i^C}.
\end{equation}
The test statistic is again the squared Mahalanobis distance $d(\bs{\hat\tau_{kj}^i}, \Theta_{2,kj}^i;\bs{\Sigma_{kj}^i})^2$ between $\bs{\hat{\tau}_{kj}^i}$ and $\Theta_{2,kj}^i$, which depends on the $(2,2)$ covariance matrix $\bs{\Sigma_{kj}^i}$ associated to $\hat{\tau}_{ji},\hat{\tau}_{ki}.$

As for the test $G=1$, we need to find the distribution of the statistic under the null hypothesis, considering the worst situation that is when $\bs{{\tau}_{kj}^i}$ lies at the boundaries of the feasible set. However, even in such a case, the distribution of the squared Mahalanobis distance is not univocally determined. Consider the following two extreme cases (see \cite{self1987asymptotic,Chernoff1954} for details): 
\begin{enumerate}[i.]
\item the true TDOA vector $\bs{\tau_{kj}^i}$ is an interior point of an edge of the feasible set. If the size of the noise is low with respect to the length of the edge, the feasible set can be approximated with a half plane. Therefore:
$$
d(\bs{\hat\tau_{kj}^i}, \Theta_{2,kj}^i;\bs{\Sigma_{kj}^i})^2 \sim
\frac{1}{2}\chi^2_0  + \frac{1}{2}\chi^2_1;
$$
\item the true TDOA vector $\bs{\tau_{kj}^i}$ corresponds to a vertex of $\Theta_{2,kj}^i$. In this case the distribution is a mixture between a Dirac delta in zero, a chi-squared with one degree of freedom and a chi-squared with two degrees of freedom, whose mixing probabilities depend on the local geometry of the feasible set. For instance, if $\bs{\tau_{kj}^i}$ coincides with $R^0$ in Figure \ref{fig:tauimagealigned}, we have:
$$
d(\bs{\hat\tau_{kj}^i}, \Theta_{2,kj}^i;\bs{\Sigma_{kj}^i})^2 \sim
\frac{1}{4}\chi^2_0  + \frac{1}{2}\chi^2_1 + \frac{1}{4}\chi^2_2.
$$
\end{enumerate}
In general situation, the distribution of the test statistic is
$$
d(\bs{\hat\tau_{kj}^i}, \Theta_{2,kj}^i;\bs{\Sigma_{kj}^i})^2 \sim \beta_0 \chi^2_0  + \beta_1 \chi^2_1 + (1-\beta_0-\beta_1)\chi^2_2,
$$
with mixing probabilities $(\beta_0,\beta_1,1-\beta_0-\beta_1)$ depending on the location of $\bs{\tau_{kj}^i}\,.$

This means that also the $p$-value of the test is a function of the true TDOAs. However, following usual statistical practice \cite{Chernoff1954} (and also for containing the computational cost of the algorithm), we are interested in defining a test that does not depend on the unknown true parameters of the model. This is achieved by globally approximating $\beta_0,\beta_1$ with the values that best describes the distribution of the squared distance in most situations. We remind that we are assuming the absence of any information on the true TDOAs and the hypothesis of low noise on the measurements, compared to the distances between the sensors. Therefore, it is justified to deduce that the generic $\bs{\tau_{kj}^i}$ belonging to the boundary of $\Theta_{2,kj}^i$ is not too close (in terms of Mahalanobis distance) to any vertex of the feasible set. Based on the previous discussion, we consequently choose $\beta_0=\beta_1=\frac{1}{2}$ and so the $p$-value of test \eqref{eq:test_couple} is:
\begin{equation}\label{eq:pvalue_couple}
	{\lambda}_{kj}^i = \frac{1}{2}\left( 1-F_{\chi^2_1}( d(\bs{\hat\tau_{kj}^i},\Theta_{2,kj}^i;\bs{\Sigma_{kj}^i})^2) \right).
\end{equation}
If ${\lambda}_{kj}^i \leq \alpha$, there is statistical evidence to say that the vector $\bs{\tau_{kj}^i}$ has at least one outlier component. In this approach, the acceptance region for $G=2$ test is the offset of the feasible set $\Theta_{2,kj}^i$ at Mahalanobis distance
$\sqrt{F_{\chi^2_1}^{-1}(1-2\alpha)}$.



In Section~\ref{sec:outlier_removal}, we will show how to merge the test results on the various pairs of TDOAs in order to detect outlier measurements. In the following, we focus instead on the efficient computation of $d(\bs{\hat\tau_{kj}^i},\Theta_{2,kj}^i;\bs{\Sigma_{kj}^i}).$ If $\bs{\hat{\tau}_{kj}^i}$ is feasible, then the distance is equal to zero. Otherwise, we should take the distance of $\bs{\hat{\tau}_{kj}^i}$ from the topological boundary $\partial\Theta_{2,kj}^i$ of the feasible set. Our strategy is based on two points:
\begin{itemize}
\item
thanks to the preprocessing performed on individual TDOAs and detailed in Section \ref{subsec:k_1}, we assume that $\bs{\hat{\tau}_{kj}^i}$ lies in the rectangle $\mathcal{I}_{ji}\times \mathcal{I}_{ki};$
\item
for every configuration of the sensors, we define the \emph{simplified acceptance region} of the measurements, which is a good approximation of the true acceptance region with more manageable geometric features.
\end{itemize}

There are two different situations, depending on the sensors arrangement.
\subsubsection{Sensors in general position} 
\begin{figure}[htb]
\begin{center}
\resizebox{4.2cm}{!}{
  \includegraphics{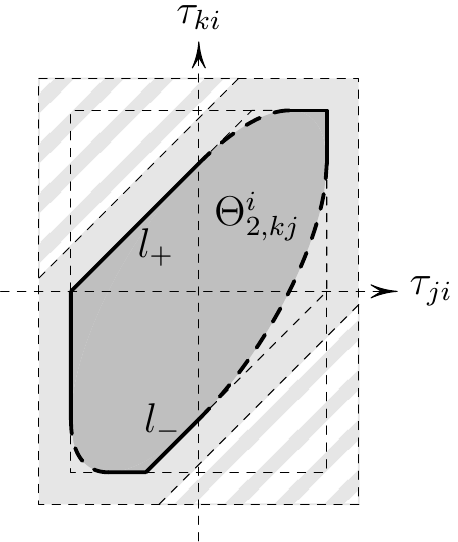}}
  \caption{\label{fig:tauimageout}
Given $\m{i}=(0,0,0),\m{j}=(0,1,0),\m{k}=(1,1,0),$ in medium gray we depict $\Theta_{2,kj}^i.$ Once the analysis for $G=1$ has been accomplished, $\bs{\hat{\tau}_{kj}^i}$ is guaranteed to be in the rectangle $\mathcal{I}_{ji}\times \mathcal{I}_{ki}$ delimited by the external dashed lines. The only unexploited inequalities are the ones related to the lines $l_{\pm}.$ Here, we aim at removing the TDOAs in the dashed gray region, so that the simplified acceptance region is the hexagon given by the union of the medium and light gray sets.}
\end{center}
\end{figure}

$\partial\Theta_{2,kj}^i$ is a complicated set, being the union of three arcs of ellipse and six linear segments (see Figure \ref{fig:tauimageout}).
In order to compute the exact distance we should use an ad hoc algorithm (see \cite{Compagnoni2016c}). Since this computation is to be performed for every $\bs{\hat{\tau}_{kj}^i},$ this could have a significant impact on the computational costs of the outlier removal procedure. For this reason, let us address an easier problem.

The idea is to approximate the feasible set $\Theta_{2,kj}^i$ with the hexagon $P_{2,kj}^i$ defined by:
\begin{subequations}\label{eq:politopoijk}
\begin{align}[left = \empheqlbrace\,]
& -d_{ji} \leq \tau_{ji} \leq d_{ji}\label{eq:politopoijk1}\\
& -d_{ki} \leq \tau_{ki} \leq d_{ki}\label{eq:politopoijk2}\\
& -d_{kj} \leq \tau_{ki} - \tau_{ji} \leq d_{kj}\label{eq:politopoijk3}
\end{align}
\end{subequations}
Thus, the goal becomes the elimination of the points $\bs{\hat\tau_{kj}^i}$ that are sufficiently far from the hexagon.

The first four inequalities (\ref{eq:politopoijk1},\ref{eq:politopoijk2}) have been exploited in the analysis for $G=1$ and the outliers that do not satisfy them have already been removed. Hence, the only inequalities to take care are \eqref{eq:politopoijk3}. Let $l_\pm$ be the lines defined by $\tau_{ki} - \tau_{ji} = \pm d_{kj}$, respectively, and $\mb{n}=(1\ -1)^T$. Through straightforward computations, we obtain the Mahalanobis distances between $\bs{\hat{\tau}_{kj}^i}$ and the lines $l_\pm$:
$$
d(\bs{\hat{\tau}_{kj}^i},l_\pm;\bs{\Sigma_{kj}^i})=\frac{\vert\hat\tau_{ki} - \hat\tau_{ji} \mp d_{kj}\vert}{\Vert\mb{n}\Vert_{\bs{\Sigma_{kj}^i}}}\,.
$$
Finally, the approximation of $d(\bs{\hat\tau_{kj}^i},\Theta_{2,kj}^i;\bs{\Sigma_{kj}^i})$ is given by
\begin{equation}\label{eq:2Ddist}
\!\!f(\bs{\hat\tau_{kj}^i};\bs{\Sigma_{kj}^i})\middleequal
\left\{\!\!\!\begin{array}{ll}
0 & \!\!\text{if}\ |\hat\tau_{ki} - \hat\tau_{ji}| \leq d_{kj},\\[1mm]
d(\bs{\hat{\tau}_{kj}^i},l_+;\bs{\Sigma_{kj}^i}) & \!\!\text{if}\ \hat\tau_{ki} - \hat\tau_{ji} > d_{kj},\\[1mm]
d(\bs{\hat{\tau}_{kj}^i},l_-;\bs{\Sigma_{kj}^i}) &  \!\!\text{if}\ \hat\tau_{ki} - \hat\tau_{ji} < -d_{kj}.
\end{array}\right.
\end{equation}
From a geometric standpoint, this function gives the distance between $\bs{\hat{\tau}_{kj}^i}$ and the strip bounded by $l_\pm.$

The simplified acceptance region associated to $f(\bs{\hat\tau_{kj}^i};\bs{\Sigma_{kj}^i})$  is depicted in Figure \ref{fig:tauimageout}. The approximation in it concerns the three dashed curvilinear boundaries of $\Theta_{2,kj}^i$ and the vertexes $R^0,R^1,R^2,$ where the true acceptance set has rounded edges.

\subsubsection{Sensors in aligned position} the topological closure of $\Theta_{2,kj}^i$ is the triangle defined by
\begin{subequations}\label{eq:parallelogrammaijk}
\begin{align}[left = \empheqlbrace\,]
& \,c_{kji}^{--+}\,\tau_{ji}+2\,d_{ji}\,\tau_{ki}\leq d_{ji}\,c_{kji}^{-++}\label{eq:parallelogrammaijk1}\\
& \,2\,d_{ki}\,\tau_{ji}+c_{kji}^{--+}\,\tau_{ki}\leq d_{ki}\,c_{kji}^{+-+}\label{eq:parallelogrammaijk2}\\
& \,c_{kji}^{+--}\,\tau_{ji}+c_{kji}^{-+-}\,\tau_{ki}\leq d_{kj}\,c_{kji}^{++-}\label{eq:parallelogrammaijk3}
\end{align}
\end{subequations}
where $c^{\pm\pm\pm}_{kji}=\pm d_{ji}\pm d_{ki}\pm d_{kj}.$
Let us name $l_{ji},l_{ki},l_{kj}$ the lines containing $\{R_i,R_j\},\{R_i,R_k\},\{R_j,R_k\},$ respectively. They support the boundary of $\Theta_{2,kj}^i$.
The corresponding distances of a point $\bs{\hat{\tau}_{kj}^i}$ from the lines are:
$$
d(\bs{\hat{\tau}_{kj}^i},l_{ji};\bs{\Sigma_{kj}^i})=
\frac{\vert c_{kji}^{--+}\,\hat\tau_{ji}+2\,d_{ji}\,\hat\tau_{ki}-d_{ji}\,c_{kji}^{-++}\vert}
{\Vert(c_{kji}^{--+}\quad 2\,d_{ji})^T\Vert_{\bs{\Sigma_{kj}^i}}}\,,
$$
$$
d(\bs{\hat{\tau}_{kj}^i},l_{ki};\bs{\Sigma_{kj}^i})=
\frac{\vert 2\,d_{ki}\,\hat\tau_{ji}+c_{kji}^{--+}\,\hat\tau_{ki}-d_{ki}\,c_{kji}^{+-+}\vert}
{\Vert(2\,d_{ki}\quad c_{kji}^{--+})^T\Vert_{\bs{\Sigma_{kj}^i}}}\,,
$$
$$
d(\bs{\hat{\tau}_{kj}^i},l_{kj};\bs{\Sigma_{kj}^i})=
\frac{\vert c_{kji}^{+--}\,\hat\tau_{ji}+c_{kji}^{-+-}\,\hat\tau_{ki}-d_{kj}\,c_{kji}^{++-}\vert}
{\Vert(c_{kji}^{+--}\quad c_{kji}^{-+-})^T\Vert_{\bs{\Sigma_{kj}^i}}}\,.
$$

According to the configuration of sensors, one or two inequalities in \eqref{eq:parallelogrammaijk} correspond to the triangular inequalities for the single TDOAs (see Figure \ref{fig:tauimagealout}). Thus, they have already been used in the analysis for $G=1$. We have the following situations:
\begin{figure}[t]
\begin{center}
\begin{minipage}[c]{.24\textwidth}
\resizebox{4.4cm}{!}{
  \includegraphics{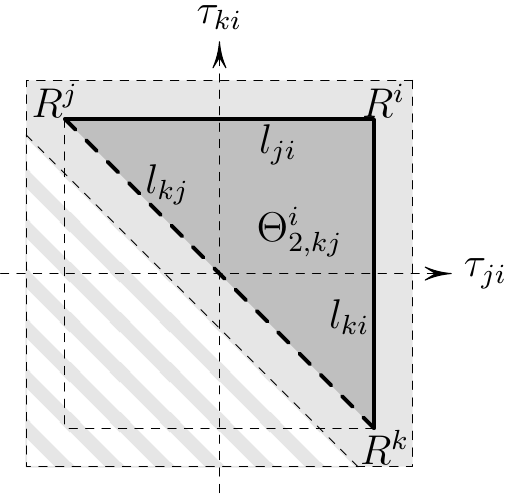}}
\end{minipage}\hspace{0mm}
\begin{minipage}[c]{.24\textwidth}
\resizebox{4.4cm}{!}{
  \includegraphics{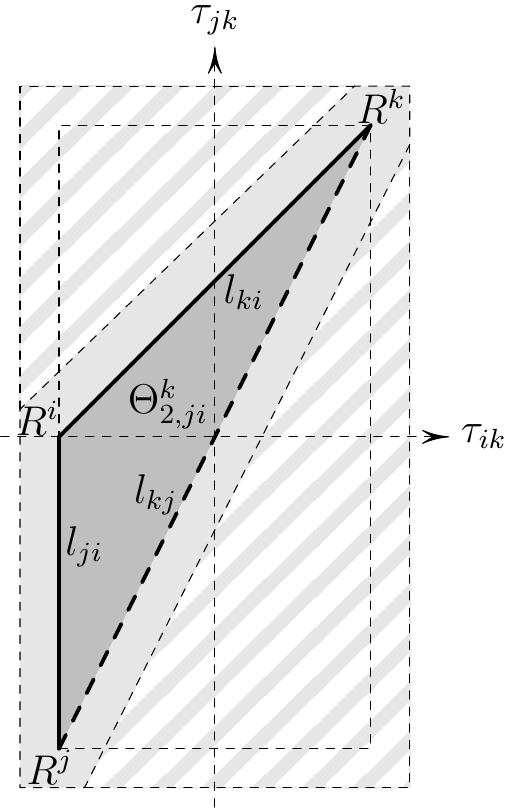}}
\end{minipage}
  \caption{\label{fig:tauimagealout}
The feasible sets $\Theta_{2,kj}^i$ and $\Theta_{2,ji}^k,$ for the sensors $\m{i}=(0,0,0),\m{j}=(1,0,0),\m{k}=(-1,0,0).$ Once the analysis for $G=1$ has been accomplished, the TDOAs are guaranteed to be in the rectangles delimited by the external dashed lines. In the $G=2$ analysis, we remove the TDOAs in the dashed gray regions. On the left, the simplified acceptance region is given by the inequality related to the diagonal $l_{kj}.$ On the right, we use the two inequalities related to the lines $l_{ki}$ and $l_{kj}.$}
\end{center}
\end{figure}
\begin{enumerate}[i.]
\item
if $\m{i}$ lies between $\m{j}$ and $\m{k},$ the feasible set is the one depicted on the left of Figure \ref{fig:tauimagealout}. The only inequality that has not been used is \eqref{eq:parallelogrammaijk3}.
In this case, the relevant function for the outlier removal procedure is
\begin{equation}\label{eq:2Ddistaligned1}
\!\!\!f(\bs{\hat\tau_{kj}^i};\bs{\Sigma_{kj}^i})\!=\!
\left\{\!\!\!\begin{array}{ll}
0 & \text{if \eqref{eq:parallelogrammaijk3} holds,}\\
d(\bs{\hat{\tau}_{kj}^i},l_{kj};\bs{\Sigma_{kj}^i}) &  \text{otherwise.}
\end{array}\right.
\end{equation}
It gives the distance between $\bs{\hat{\tau}_{kj}^i}$ and the half--plane bounded by $l_{jk}$ and containing $\Theta_{2,kj}^i$;
\item
if $\m{i}$ does not lie between $\m{j}$ and $\m{k},$ we obtain a feasible set of the type depicted on the right of Figure \ref{fig:tauimagealout}. The relevant inequalities are \eqref{eq:parallelogrammaijk1} and \eqref{eq:parallelogrammaijk2}, that define four distinct regions of the real plane. However, we are interested in the TDOAs lying on the rectangle $\mathcal{I}_{ji}\times \mathcal{I}_{ki}.$ Therefore, in this case we define the function
\begin{equation}\label{eq:2Ddistaligned2}
\!\!\!\!f(\!\bs{\hat\tau_{kj}^i};\!\bs{\Sigma_{kj}^i}\!)\middleequal\!
\left\{\!\!\!\!\begin{array}{ll}
d(\!\bs{\hat{\tau}_{kj}^i},l_{ki};\bs{\Sigma_{kj}^i}\!) &\!\!\!  \text{if only (\ref{eq:parallelogrammaijk1}) holds,}\\
d(\!\bs{\hat{\tau}_{kj}^i},l_{ji};\bs{\Sigma_{kj}^i}\!) &\!\!\! \text{if only (\ref{eq:parallelogrammaijk2}) holds,}\\
0 &\!\!\!  \text{otherwise.}
\end{array}\right.
\end{equation}
Geometrically, it gives the distance between $\bs{\hat{\tau}_{kj}^i}$ and the line supporting the closest diagonal side of $\Theta_{2,kj}^i$, for points $\bs{\hat{\tau}_{kj}^i}$ in $\mathcal{I}_{ji}\times \mathcal{I}_{ki}$ but not belonging to the feasible set (see Figure \ref{fig:tauimagealout}).
\end{enumerate}

\vspace{1mm}
The simplified acceptance regions for the two scenarios are depicted in Figure \ref{fig:tauimagealout}. Again, we have an approximation near the vertexes $R^0,R^1,R^2,$ where the true acceptance set has rounded edges. Preliminary results for sensor configurations tested in Section~\ref{sec:evaluation} showed that the functions $f(\bs{\hat\tau_{kj}^i};\bs{\Sigma_{kj}^i})$ defined in \eqref{eq:2Ddist}, \eqref{eq:2Ddistaligned1} and \eqref{eq:2Ddistaligned2} well approximate  $d(\bs{\hat\tau_{kj}^i},\Theta_{2,kj}^i;\bs{\Sigma_{kj}^i})$ in the context of outlier removal. We can therefore run the statistical tests using an approximation of the $p$-values \eqref{eq:pvalue_couple} defined as
\begin{equation}\label{eq:approx_pvalue_couple}
{\lambda}_{kj}^i \approx\frac{1}{2}\left( 1-F_{\chi^2_1}( f(\bs{\hat\tau_{	kj}^i};\bs{\Sigma_{kj}^i})^2) \right).
\end{equation}
We postpone to future investigations a deeper study on the effect of this choice, considering the impact of different array geometries.

\subsection{$G=3$: Test on triplets of TDOAs}

\begin{figure}[tb]
\begin{center}
\includegraphics[width=0.7\columnwidth,trim={0 2.5cm 0 3.5cm},clip]{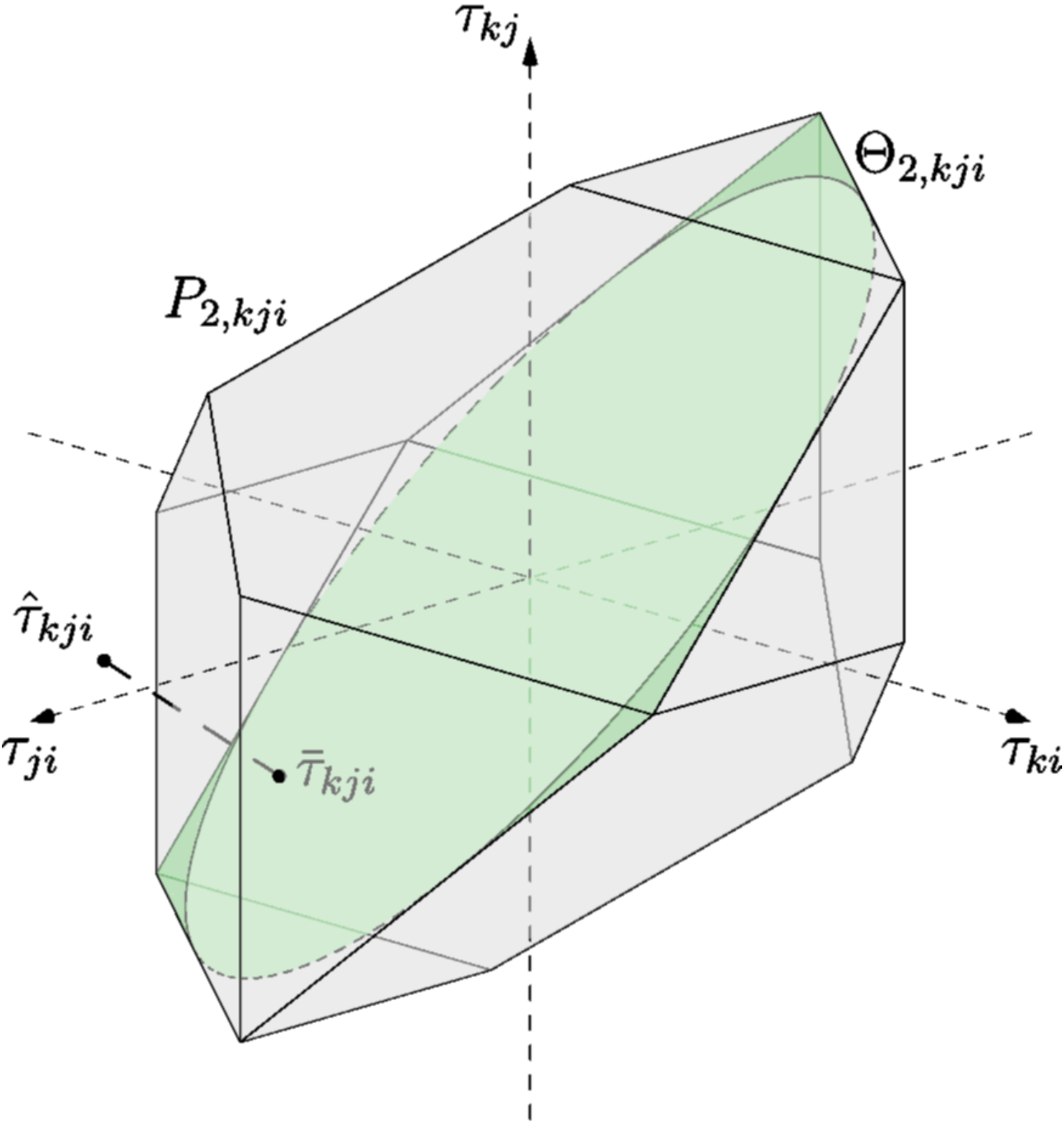}
\caption{\label{fig:taucomimageout} The feasible set $\Theta_{2,kji}$ is a subset of the polytope $P_{2,kji}.$ The $G=3$ test takes care of the points $\bs{\hat\tau_{kji}}$ too far from the plane $V_{2,kji}$ containing $\Theta_{2,kji}.$}
\end{center}
\end{figure}

Consider now the triples of TDOAs estimated from the signals measured by three sensors. We call them $\bs{\hat\tau_{kji}}=(\hat\tau_{ji},\hat\tau_{ki},\hat\tau_{kj})^T\in\RR^3,$ where $0\leq i<j<k\leq n$. The feasible set of $\bs{\hat\tau_{kji}}$ is $\Theta_{2,kji}$ (see Figure \ref{fig:taucomimageout}), which is a subset of the linear space $V_{2,kji}$ defined by the zero-sum condition
\begin{equation}
\tau_{ji}-\tau_{ki}+\tau_{kj}=0.
\end{equation}
As above, we need a statistical test on the hypotheses:
\begin{equation}\label{eq:test_triple}
\!\!\!H_{0,kji}\! : \bs{\hat{\tau}_{kji}} \text{ does not contain outliers; } H_{1,kji} = H_{0,kji}^C.
\end{equation}

$\Theta_{2,kji}$ is bounded by the triangular inequalities involving $\tau_{ji},\ \tau_{ki}$ and $\tau_{kj}.$ For example, if the receivers are in general positions, we have twelve non-redundant inequalities of type \eqref{eq:politopoijk}. They define a three--dimensional polytope $P_{2,kji}$ with twelve facets and $\Theta_{2,kji}\subset P_{2,kji}.$ The goal of $G=1,2$ tests is to eliminate the outliers too far from $P_{2,kji}$. What remains to take care of are therefore the points $\bs{\hat\tau_{kji}}$ that are close to the polytope, but are too distant from $V_{2,kji}.$

Given $\mb{n}=(1\ -1\ 1)^T,$ the Mahalanobis distance between $\bs{\hat{\tau}_{kji}}$ and $V_{2,kji}$ is
\begin{equation}
d(\bs{\hat{\tau}_{kji}},V_{2,kji};\bs{\Sigma_{kji}})=\frac{\vert\hat\tau_{ji} - \hat\tau_{ki} + \tau_{kj}\vert}{\Vert\mb{n}\Vert_{\bs{\Sigma_{kji}}}}\,,
\end{equation}
where $\bs{\Sigma_{kji}}$ is the covariance matrix associated to the three TDOAs that we are considering. In this case, if $\bs{\hat{\tau}_{kji}}$ does not contain outliers, one has exactly
\begin{equation}\label{eq:chisq_3D}
d(\bs{\hat{\tau}_{kji}},V_{2,kji};\bs{\Sigma_{kji}})^2 \sim \chi^2_1.
\end{equation}
In particular, the distribution does not depend on the unknown true TDOAs.
The  $p$-value of test \eqref{eq:test_triple} is
\begin{equation}
	\lambda_{kji} = 1-F_{\chi^2_1}(d(\bs{\hat{\tau}_{kji}},V_{2,kji};\bs{\Sigma_{kji}})^2).
\end{equation}

\subsection{Discussion}

In the above analysis, we considered only some kind of subsets of the complete set of TDOA measurements. In particular, we never took $G$--tuples of TDOAs involving more than three receivers. The practical reason behind this choice is the lack of the description of the feasible sets $\Theta_n$ for $n>2.$ However, there are also statistical and computational arguments in support of our procedure. 

From a statistical standpoint, we can list two relevant advantages of considering groups of few TDOAs at a time:
\begin{itemize}
\item
in order to simplify the identification of the outliers, we need subsets of TDOAs containing few of them, ideally no more than one outlier per subset. But the number of outliers is positively correlated to the cardinality of such subsets and so it is correct to consider minimal data groups;
\item
the presence of noise on the measurements tends to mask the presence of the outliers. This is particularly true if we combine the errors from many measurements. Also in this case, by taking minimal data subsets we can contain this phenomenon most effectively.
\end{itemize}

From a computational standpoint, we should stress that our approach is based on the linear relations (i.e. the ZSCs and the triangular inequalities) that exist between the TDOAs, for $n\leq 2$. This implies that we only deal with simple closed-form expressions, which we only have to compute a limited number of times. Conversely, if $n>2$ Theorem \ref{th:Vn} tells us that, in order to work with relations independent from the ZSCs, we should consider nonlinear relations. This, however, would make the algorithms more complicated in terms of computational cost as well as robustness.

Finally, we briefly discuss the role of the squared Mahalanobis distance as test statistic. In the statistics literature, an important and widespread hypothesis test is the ratio likelihood test. Its popularity is due to the Neyman–Pearson lemma \cite{neyman1933}, which guarantees that it is the most powerful test between two simple hypothesis. In order to define this test, one need the likelihood functions of the two hypothesis. In our manuscript, we made no assumption on the distribution of the outliers. However, in the absence of any information, it is reasonable to think that an outlier $\hat\tau_{ji}$ could assume any value in a sufficiently large set containing the feasible set $\Theta_{1,ji},$ with equal probability. In this hypothesis, one can compute the log-likelihood ratio and easily obtain that it is equivalent (up to an additive constant) to the squared Mahalanobis distance of the measured data from the feasible set. This reasoning strongly supports our choice of the test statistic.

\section{Outlier identification}\label{sec:outlier_removal}

In the previous Section, we showed how to perform statistical tests on each Group of $G=1$, $G=2$ or $G=3$ TDOAs separately. In the case of $G=1$, each test directly identifies whether the considered TDOA is an outlier. For this reason, this step is treated as a pre-processing that is always applied to remove outliers before proceeding with other tests. After pre-processing, we have up to $3 \binom{n+1}{3}$ and $\binom{n+1}{3}$ (possibly dependent) tests for $G=2$ and $G=3$, respectively. These tests, performed separately, are only able to determine whether at least one TDOA in the group is a possible outlier, but do not provide any information on which one it is.

In this section, we focus on outlier identification within the available TDOAs by merging results from tests performed on multiple $G$-tuples. This is done by exploiting two methods available in the statistical literature: \textit{multiple testing} \cite{dudoit2007multiple} and \textit{combined testing} \cite{birnbaum1954combining}, \cite{loughin2004systematic}. More specifically, we devise an iterative algorithm that exploits multiple testing to detect whether outliers are still present within the available  measurements, and combined testing to remove them.



\subsection{Multiple testing}

In both cases of $G=2$ and $G=3$, each TDOA $\hat{\tau}_{ji}$ is included in several tests (namely, all triplets of sensors containing the $i$th and $j$th ones). In Section~\ref{sec:stat_models}, we showed how to control the probability of false detection in each separate test comparing $p$-values to a threshold $\alpha$. However, in order to control the whole family of tests when they are considered at once, a $p$-value correction strategy must be applied according to the dependency structure of the tests.
For example, let $H_{0,ji}^{(1)},\ldots,H_{0,ji}^{(M)}$ be a set of $M$ null hypotheses pertaining $\hat{\tau}_{ji}$, $\lambda_{ji}^{(m)}$ be the corresponding $p$-values, and $I_0$ be the set of tests corresponding to true null hypotheses. In the case of  independent tests, if we set a value $\alpha$ for the probability of false discovery of each test, the probability of ending up with at least one false discovery on the whole family of tests becomes
\begin{equation}\label{eq:exact.test.pD}
\begin{split}
&P[\text{reject at least one }H_{0,ji}^{(m)} | H_{0,ji}^{(m)}  \text{ true } \forall m \in I_0]  \\ 
= &1 - P[\text{accept all }H_{0,ji}^{(m)}  | H_{0,ji}^{(m)}  \text{ true } \forall m \in I_0]  \\ 
= &1 - P[\cap_{m \in I_0}  (\text{accept }H_{0,ji}^{(m)}  | H_{0,ji}^{(m)}  \text{ true })]  \\ 
= &1-(1-\alpha)^{M_0} > \alpha,
\end{split}
\end{equation}
where $M_0$ denotes the cardinality of the set $I_0$, i.e. the number of true null hypotheses.
This is why, when testing multiple hypotheses, it is common practice to adjust the test results in such a way to account for multiplicity. 

Several multiple testing techniques exist in the statistics literature \cite{dudoit2007multiple}. The method that we exploit in this article is the Benjamini-Hochberg (BH) adjustment \cite{Benjamini1995}, \cite{benjamini2001control}, which computes adjusted $p$-values $\lambda_{ji,\text{\tiny{BH}}}^{(m)}$ for each test as follows:
\begin{itemize}
\item order the $p$-values in increasing order so that $\lambda_{ji}^{(1)} \leq \lambda_{ji}^{(2)} \leq \ldots \leq \lambda_{ji}^{(M)}$;
\item compute the adjusted $p$-values $\lambda_{ji,\text{\tiny{BH}}}^{(m)}= \lambda_{ji}^{(m)}
{M}/{m}$.
\end{itemize}
After the multiplicity adjustment, if
\begin{equation}\label{eq:testcorrectedpvalue}
\overline{\lambda}_{ji}=\min_m  \{\lambda_{ji,\text{\tiny{BH}}}^{(m)} \; \vert \; m = 1,\ldots , M  \} \leq \alpha
\end{equation}
we say that $\hat{\tau}_{ji}$ is a potential outlier. However, the low value of $\overline{\lambda}_{ji}$ could be caused by the other TDOAs involved in the corresponding test. To overcome this ambiguity, in the next subsection we exploit combined testing.

\subsection{Combined testing}
If at least one $\overline{\lambda}_{ji}$ satifies condition \eqref{eq:testcorrectedpvalue}, we infer that the TDOAs dataset is affected by at least one outlier. In order to compute how likely it is for $\hat{\tau}_{ji}$ to be the searched outlier, all tests concerning this TDOA must to be considered and combined. This can be done by simply applying the standardized Fisher combination function \cite{fisher1925statistical}, \cite{fisher1948answer}
\begin{equation}\label{eq:Fisher_test}
T_{ji}= -\frac{2}{M}\sum_{m =1}^M \ln (\lambda_{ji}^{(m)} )\;.
\end{equation}
This statistic is a combination of all tests including $\hat{\tau}_{ji}$. The lower are the $p$-values, the greater is $T_{ji}$. By computing \eqref{eq:Fisher_test} for all the measurements, the TDOA associated to the largest $T_{ji}$ can be considered as outlier and can thus be removed from the set. 

\subsection{Algorithm}
\label{subsec:algo}

The statistical procedures described above are jointly used in our proposed outlier removal algorithm. We start from pre-processed TDOAs, which are those that fall within the acceptance region defined in \eqref{eq:outlier_interval}. For all the sensor pairs we correct the $p$-values using the BH adjustment procedure. If at least one of these $p$-values is smaller than $\alpha$, we conclude that the measurements contain outliers. If so, we compute the Fisher's statistic \eqref{eq:Fisher_test} and we remove from the measurement set the TDOA(s) associated to the greatest $T_{ji}$ value. The whole procedure is iterated until no outliers are detected anymore. The detailed procedure is shown in Algorithm~\ref{alg:tdoa_outlier_removal}.
 
This algorithm can be applied considering only $G=2$ or $G=3$, and it can also be applied first to $G=2$ and then to $G=3$ or viceversa.
\RestyleAlgo{boxruled}
\newlength\mylen
\newcommand\mydata[1]{%
	\settowidth\mylen{\KwData{}}%
	\setlength\hangindent{\mylen}%
	\hspace*{\mylen}#1\\}
\begin{algorithm}	
	\caption{TDOA outlier removal}
	\SetKwInput{Input}{Input}
	\SetKwInput{Output}{Output}
	\DontPrintSemicolon
	\BlankLine
	
	\KwData{Preprocessed TDOAs: $\mathcal{T} = \left\lbrace \hat{\tau}_{ji}\in \mathcal{I}_{ji}\, : \, j > i \right\rbrace$}
	\mydata{Sensor positions: $\mathbf{m}_i\,,i=0\ldots n$}
	\BlankLine
	\KwIn{Significance level $\alpha$ (default $\alpha=5\%$)}
	\BlankLine
	\KwOut{Outlier-free TDOAs: $\mathcal{T}_{\text{\tiny F}}$}
	\BlankLine
	
	\Begin
	{
		$\mathcal{T_{\text{\tiny F}}} \leftarrow \mathcal{T}$\\
		$\mathcal{U} \leftarrow \left\lbrace (j,i) \vert \hat{\tau}_{ji} \in \mathcal{T}\right\rbrace$\\
		$\mathcal{W} \leftarrow \begin{cases}
		\scriptstyle
		\{\bs{\hat{\tau}_{kj}^i}\,\vert\, 0\leq j<k\leq n,\, k\neq i ,\, j\neq i\}  & \scriptstyle \text{if} \; G=2 \; \text{(TDOA pairs)}  \\
		\scriptstyle
		\{\bs{\hat{\tau}_{kji}}\,\vert\, 0\leq i < j < k \leq n\}  & \scriptstyle \text{if}  \; G=3 \; \text{(TDOA triplets)}  
		\end{cases}$\\ 
		\BlankLine 
		$\textsc{exit} \leftarrow 0$ \tcp{stop condition}
		\While{$\textsc{exit}==0$}
		{
			\BlankLine
			\ForEach{$(j,i) \in \mathcal{U}$}
			{
				\BlankLine
				$\mathcal{W}_{ji} \leftarrow \left\{ \text{elements of } \mathcal{W} \text{ containing } \hat{\tau}_{ji} \right\}$\\
				$\Lambda_{ji} \leftarrow \left. \{\lambda_{ji,\text{\tiny{BH}}}^{(m)}\}\right\vert_{\scriptscriptstyle m=1}^{\scriptscriptstyle \vert\mathcal{W}_{ji}\vert} $			
				\tcp{adjusted p-values}
				$\overline{\lambda}_{ji}\leftarrow \min \Lambda_{ji} $\\
				$T_{ji} \leftarrow \frac{-2}{\vert\mathcal{W}_{ji}\vert}\sum_{\scriptscriptstyle m=1}^{\scriptscriptstyle \vert\mathcal{W}_{ji}\vert} \ln (\lambda_{ji}^{(m)})$ \tcp{Fisher stat}				
			}	
			
			\BlankLine 	
			\tcp{multiple test}				
			\uIf{$\min \{ \overline{\lambda}_{ji} \}_{\scriptscriptstyle (j,i)\in \mathcal{U}} > \alpha $}
			{
				$\textsc{exit}\leftarrow 1$\\				
			}
			\Else
			{
				\tcp{combined test}
				$(\overline j,\overline i) \leftarrow \argmax_{(j,i)}\; \{ T_{ji}\}_{\scriptscriptstyle (j,i)\in \mathcal{U}}$ \tcp{outlier idx}
				\BlankLine
				\tcp{remove outlier TDOA}
				$\mathcal{T_{\text{\tiny F}}} \leftarrow \mathcal{T_{\text{\tiny F}}} \setminus \{\hat{\tau}_{\overline j \overline i}\} $
				\BlankLine
				\tcp{update the sets}
				$\mathcal{U} \leftarrow \mathcal{U} \setminus \{(\overline j,\overline i)\}$\\
				$\mathcal{W} \leftarrow \left\{\text{elements of } \mathcal{W} \text{ not containing } \hat{\tau}_{\overline j \overline i} \right\}$\\
			}		
			
		}

	}
	\label{alg:tdoa_outlier_removal}
\end{algorithm}

\subsection{Computational complexity}
We now analyze the computational complexity of Algorithm~\ref{alg:tdoa_outlier_removal}. To do so, we evaluate the number of iterations of the two loops involved in the algorithm. The most extern \textit{while} loop is iterated until all the outliers are removed from the set of pairs $\mathcal{U}$. Thus, in the presence of $Z$ outliers, it is in principle executed for a maximum of $Z$ iterations. The innermost \textit{for} loop cycles on the elements of $\mathcal{U}$. We notice that complexity is dominated by the computation of the adjusted $p$-values, which involves the sorting of the elements of $\mathcal{W}_{ji}$. The complexity at each iteration is therefore in the order of $O( \vert \mathcal{W}_{ji}\vert \log \vert \mathcal{W}_{ji}\vert)$, where $\vert \mathcal{W}_{ji}\vert)$ denotes the cardinality of $\mathcal{W}_{ji}$.
It is worth noticing that number of elements in both $\mathcal{U}$ and $\mathcal{W}_{ji}$ diminishes at each iteration of the external loop. However, the number of elements removed at each iteration is difficult to predict, as it depends on the number of outliers. We therefore limit ourselves to provide an upper bound for the overall complexity, assuming that the cardinalities of $\mathcal{U}$ and $\mathcal{W}_{ji}$ remains constant to their maximum values, corresponding to the initial conditions. In particular, in the worst case, the cardinality of $\mathcal{U}$ corresponds to the number of microphone pairs, i.e. $\vert \mathcal{Q}\vert= q = (n+1)n/2$. The cardinality of $\mathcal{W}_{ji}$ is given by the number of elements of $\mathcal{W}$ in which the $i$th and $j$th microphones are involved. It is easy to verify that $\vert \mathcal{W}_{ji}\vert = 2(n-1)$ when $G=2$ and $\vert \mathcal{W}_{ji}\vert = n-1$ when $G=3$. Therefore, in general $\vert \mathcal{W}_{ji}\vert \propto n$.

In the light of these observations, we can conclude that the upper bound for the computational complexity is
\begin{equation}
O(Z\cdot q \cdot n \log(n))\;.
\end{equation}
Expressing it with respect to the number of microphones, we finally obtain
\begin{equation}
O(Z \cdot n^3 \log(n))\;.
\end{equation}
In the worst possible scenario, the number of outliers approaches the number of measurements, i.e. $Z\approx q$, thus the upper bound rises to $O(n^5\log(n))$ in this case. However, in practical situations it is more realistic to assume $Z\ll q$.

\subsection{Remarks}\label{sec:remarks}

In the described Algorithm the BH adjustment is separately performed on each set of $p$-values pertaining to each TDOA $\hat{\tau}_{ji}$ separately. This choice favors the possibility of discarding some inliers rather than missing possible outliers. This is the most natural choice for all the applications in which even a few outliers could severely hinder the achieved result (e.g., source localization, array calibration, etc.).

For those applications where it is more important to preserve the largest number of inliers even at the price of a larger number of outliers, a more conservative procedure can be devised. Indeed, it would be possible to adjust in a single step all $p$-values of all TDOAs. This variant would lead to a more conservative procedure, the study of which is beyond the scope of this paper.

\section{Evaluation on synthetic data}
\label{sec:evaluation}

In order to evaluate the proposed methodology for outlier removal, we run a series of tests simulating the use case of acoustic TDOA measurements with microphone arrays. Notice that the outlier removal algorithm is not strictly tailored to this specific scenario, but can be used for whatever task is based on time or range differences of arrivals.

We tested the proposed algorithm for the cases involving multiple/combined testing, i.e. for tuples of $G=2$ and $G=3$ TDOAs. The case $G=1$ was not explicitly tested, as its behavior is completely deterministic. Indeed, all measurements outside the acceptance region defined by \eqref{eq:outlier_interval} are detected as outliers with probability $1$. 
In the light of these considerations, we tested four different possibilities for detecting and removing outliers:
\begin{itemize}
	\item \texttt{G2}: Alg. \ref{alg:tdoa_outlier_removal} applied to pairs of TDOAs;  
	\item \texttt{G3}: Alg. \ref{alg:tdoa_outlier_removal} applied to triplets of TDOAs;  
	\item \texttt{G2}+\texttt{G3}: $\texttt{G2}$ followed by $\texttt{G3}$;
	\item \texttt{G3}+\texttt{G2}: $\texttt{G3}$ followed by $\texttt{G2}$.
\end{itemize}

As microphones configurations we considered: i) an uniform linear array composed by 7 microphones spaced by $10\,\mathrm{cm}$; ii) a 3D cross array composed by 7 microphones in positions $(0,0,0)^T$, $(\pm0.3, 0, 0)^T$, $(0, \pm0.3, 0)^T$, $(0, 0, \pm0.3)^T\, \mathrm{m}$. For each configuration we simulated TDOA acquisition from an acoustic source located at $\mathbf{x}$. More specifically, we corrupted the nominal TDOAs adding zero mean gaussian noise with standard deviation $\sigma=0.7\,\mathrm{cm}$ and simulating the presence of a given number $Z$ of outliers. We run Monte-Carlo simulations of this experiment testing different source positions and number of outliers. In particular 100 source positions were randomly selected within $2\,\mathrm{m}$ from the array center, considering a 2D scenario for the linear array and a 3D scenario for the cross array. Outliers were modeled by selecting $Z$ TDOAs and randomly replacing each one of them with a value uniformly distributed within the interval $$\mathcal{I}_{ji} \setminus [\tau_{ji}(\mathbf{x}) - \gamma_{\alpha,ji}, \; \tau_{ji}(\mathbf{x}) + \gamma_{\alpha,ji}]\,.$$ This means that we consider as outliers values within the acceptance region \eqref{eq:outlier_interval}, but out of $(1-\alpha)$th percentile of the Gaussian distribution describing additive noise on the nominal TDOA $\tau_{ji}(\mathbf{x})$.

As a matter of comparison, the state-of-the-art robust denoising (\texttt{R-DEN}) algorithm proposed in \cite{Velasco16} were included in the testing campaign. To enable a fair comparison, we informed this method with the exact number of outliers present for each Monte-Carlo run. Indeed, \texttt{R-DEN} require knowledge about the maximum number $\kappa$ of outliers present in the measurement set.

In order to evaluate the performance of each method, we resorted to classical statistical metrics: i) true positive rate (TPR) is the percentage of outliers TDOAs detected as such; ii) true negative rate (TNR) is as the percentage of inlier TDOAs detected as such. The larger the TPR, the larger the number of outliers we are correctly detecting. The larger the TNR, the larger the number of inliers that we are keeping. 

\begin{figure}[t]
	\centering
	\subfloat[Linear - TPR]{\includegraphics[width=0.45\columnwidth]{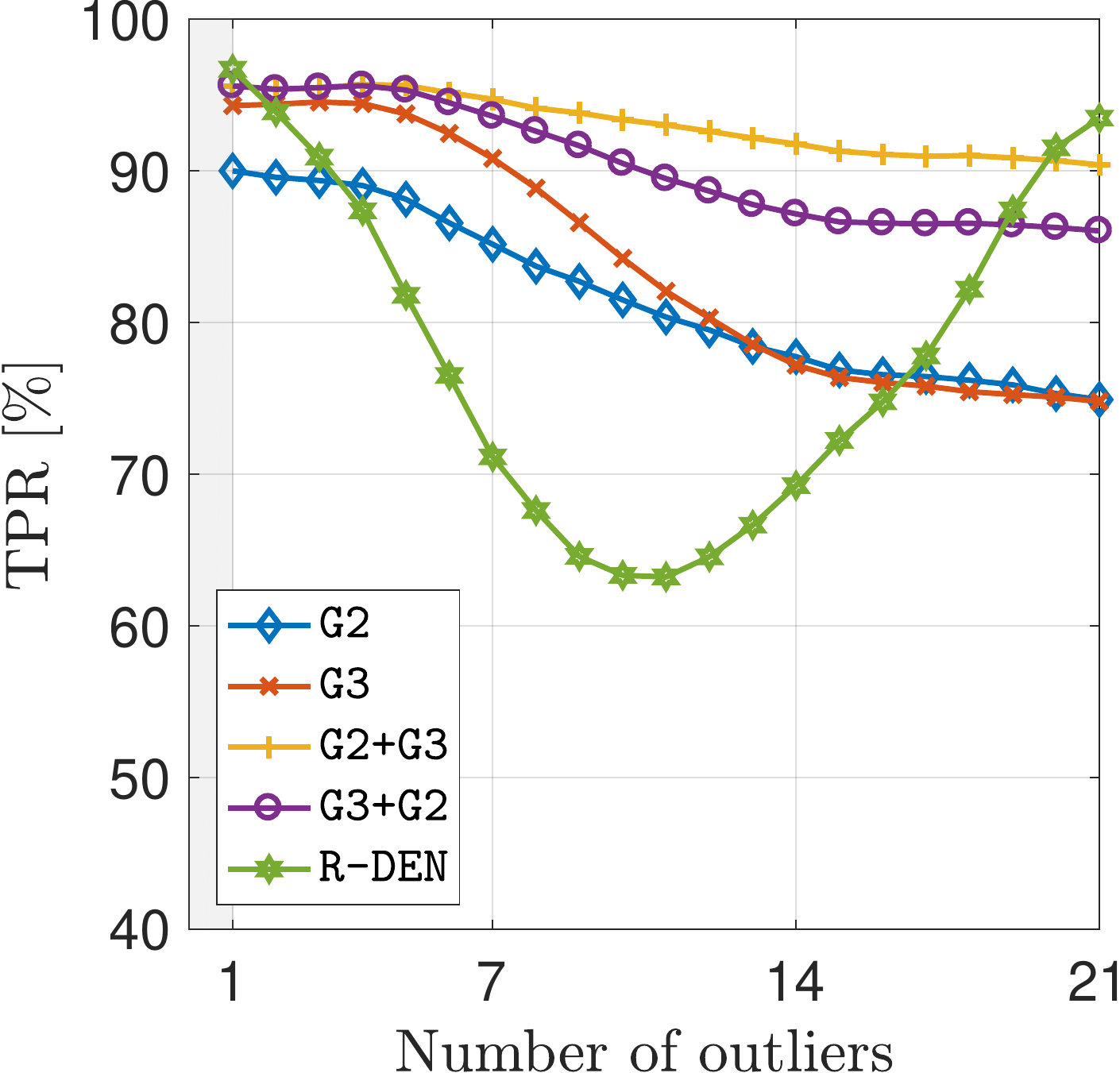}} \hfil
	\subfloat[Cross - TPR]{\includegraphics[width=0.45\columnwidth]{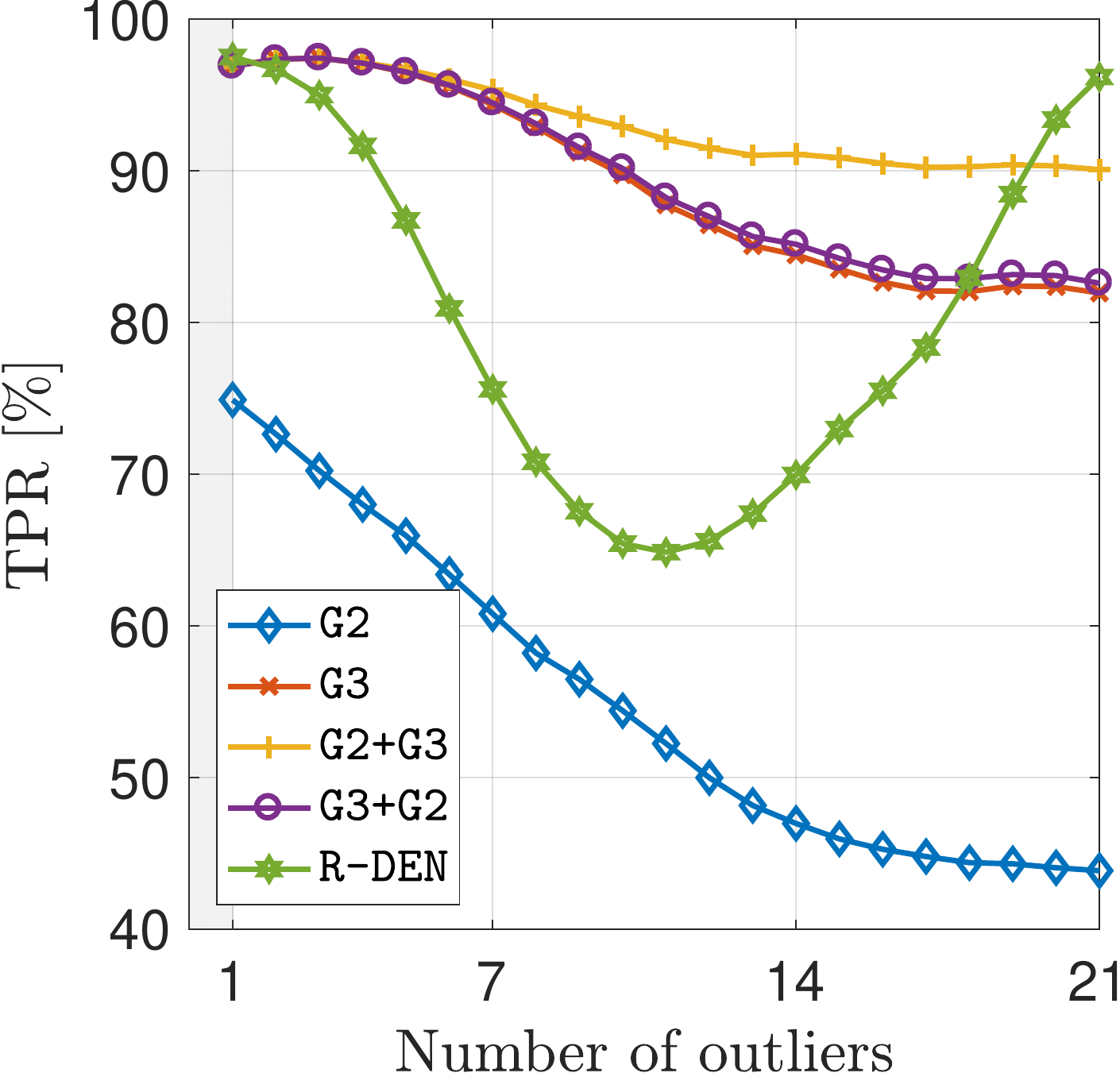}}
	\vfill
	\subfloat[Linear - TNR]{\includegraphics[width=0.45\columnwidth]{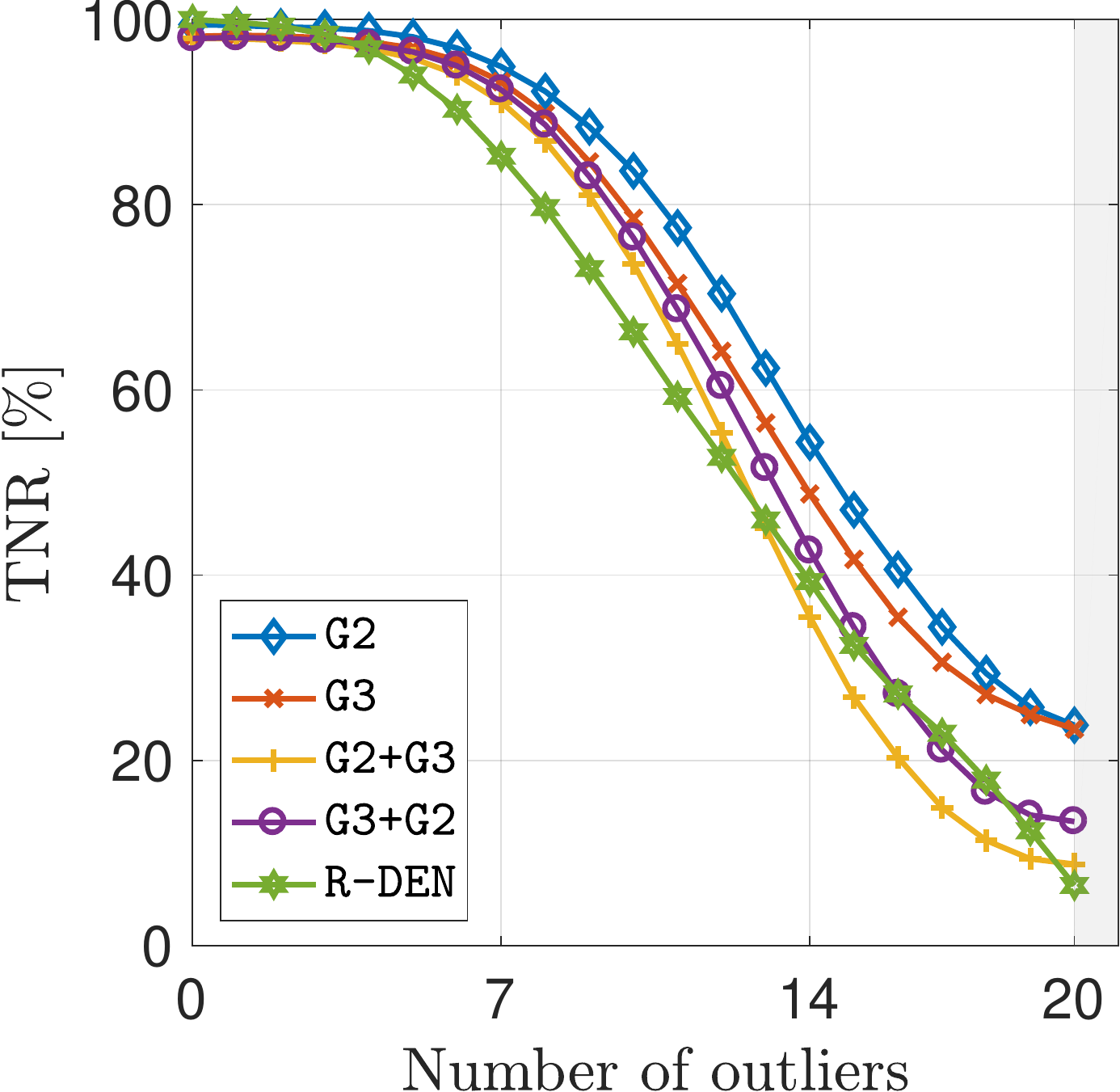}} \hfil
	\subfloat[Cross - TNR]{\includegraphics[width=0.45\columnwidth]{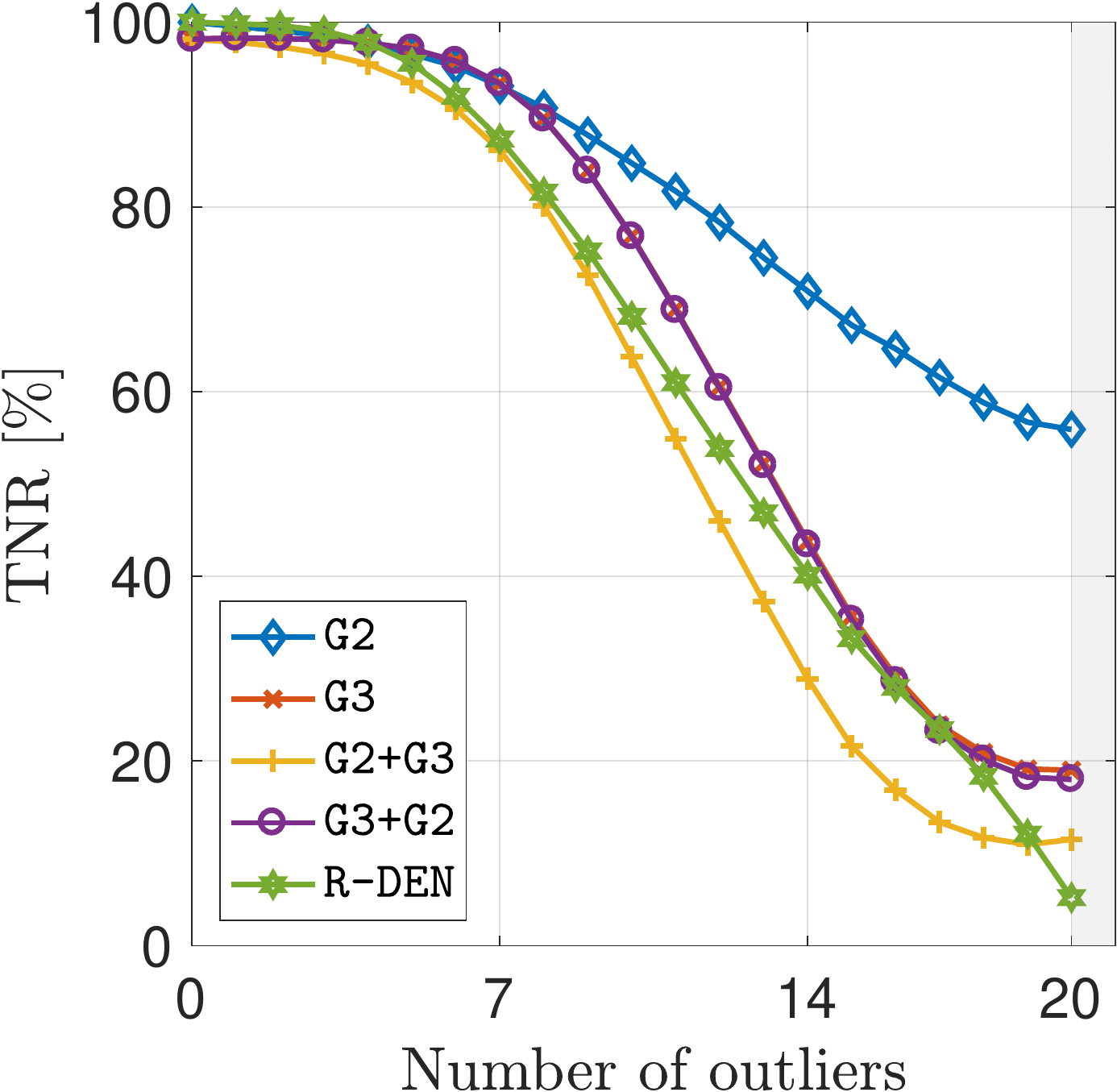}}	
	\caption{Simulation results averaged over 100 realizations using either the linear (a)(c) or cross (b)(d) array.}
	\label{fig:simulations}
\end{figure}

Figure~\ref{fig:simulations} shows the obtained results averaging 1000 Monte-Carlo runs for each source position and number of outliers. Outliers range between $Z=0$ (i.e., no outliers) and $Z=21$ (i.e., no inliers) and $\alpha=0.05$ for all methods. 
Already at a first glance it is possible to notice the inherent trade-off between TPR and TNR: methods achieving a high TPR tend to exhibit a low TNR. This means that the larger the number of outliers that we remove, the larger the number of inliers that we remove as well. Nonetheless, it is possible to achieve more promising performance, such as TPR larger than 95\% and TNR reaching nearly 100\% when 5 measurements out of 21 are outliers. 

It is interesting to notice that \texttt{G2} exhibits better performance for the linear array rather than for the cross-shaped one. As a matter of fact, in the linear case, microphones are always aligned, thus the feasible set is the one depicted in Figure~\ref{fig:tauimagealout}. Conversely, using the cross-shaped array, when microphones are not aligned the feasible set is the one depicted in Figure~\ref{fig:tauimageout}. With a finer analysis we found that in this case of non-aligned microphones the likelihood that an outlier will fall within the 2D feasible region is larger than in the case of aligned microphones. The intuition behind this behaviour is that the ratio between the dashed gray region and the feasible set area in Figure~\ref{fig:tauimagealout} is larger than in Figure~\ref{fig:tauimageout}.

From a general point of view, the use of \texttt{G2} alone leads to the worst performance in terms of TPR. However, its use together with \texttt{G3} allows us to achieve a larger TPR than using \texttt{G3} alone, which motivates its use. We also notice that \texttt{G2+G3} or \texttt{G3+G2} are the most robust choices whenever it is acceptable to throw away some 
good measurements in order to remove the largest number of outliers.

As far as \texttt{R-DEN} is concerned, we observe that it generally achieves lower TPR and TNR scores than the proposed methods. Only in case of using the cross-shaped array \texttt{R-DEN} achieves slightly superior performances compared to those of \texttt{G2}. It is worth noticing, however, that the TPR behavior of \texttt{R-DEN} is actually meaningful only when the number of outliers is lower than 10. Above this value, the TPR starts increasing since \texttt{R-DEN} tends to eliminate $\kappa$ TDOAs from the measurement set. Clearly, as the number of outliers increases approaching the total number of TDOAs, the probability of correctly identifying the outliers increases. In the extreme situation in which exactly 21 outliers are present, \texttt{R-DEN} removes almost all the measurements from the set, leading to a TPR very close to $100\%$.

\begin{figure}[t]
	\centering	
	\subfloat[Linear - ME]{\includegraphics[width=0.45\columnwidth]{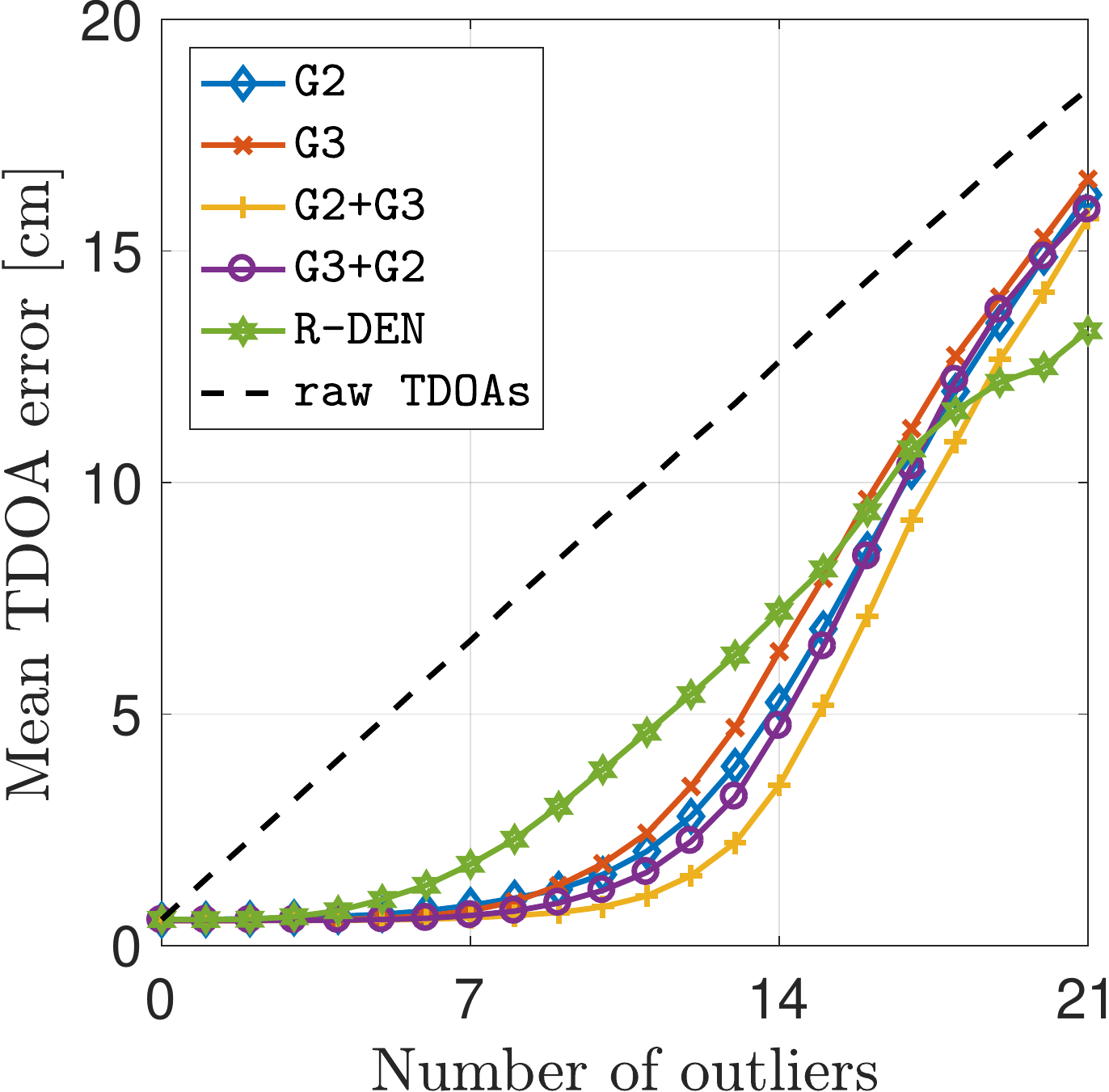}} \hfil
	\subfloat[Cross - ME]{\includegraphics[width=0.45\columnwidth]{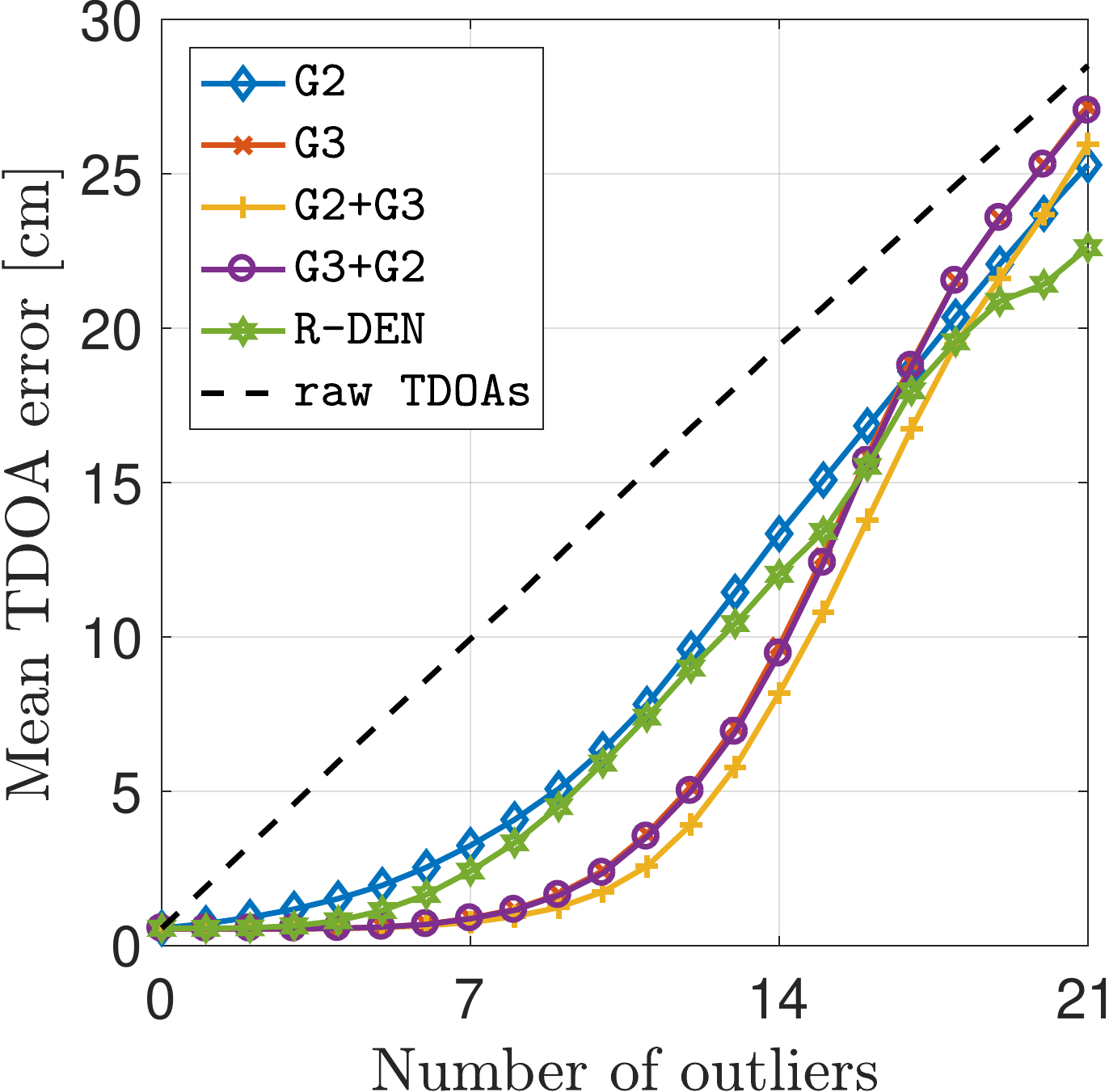}}
	\caption{Mean TDOAs error for the linear (a) and the cross (b) array. Results are compared with the mean error relative to raw TDOAs (dashed line).}
	\label{fig:results_sim_SNR}
\end{figure}

In order to give more insight about the residual error on TDOAs left after outlier removal, we compute the Mean Error
\begin{equation}
\label{eq:tdoa_snr}
\text{ME} = \frac{1}{{\vert  \mathcal{\scriptstyle U} \vert }} \sum_{(i,j)\in \mathcal{U}} {\vert \tau_{ji} - \hat{\tau}_{ji} \vert} \;,
\end{equation} 
where $\mathcal{U}$ is the set of inlier microphone pairs indexes.
Results are reported in Figure~\ref{fig:results_sim_SNR}, which shows the mean error for all the algorithms under analysis and on raw TDOAs (i.e., the whole set of measurement before outlier removal). 
We first notice that outlier removal always reduces the mean error, which means that the selected inlier TDOAs are closer to the groundtruth than the raw TDOA set. Indeed, while the mean error of raw data linearly grows with the number of outlier, outlier removal algorithms keep the mean error at bay. Moreover, the mean error of the $4$ tested algorithms confirms the general trend observed from the analysis of TPR and TNR. It is worth noticing that, using any combination of algorithms involving \texttt{G3}, it is possibile to maintain the mean error constant at the minimum level up to the presence of $10$ outliers. As expected, the \texttt{G2} performances are slightly worse for the cross array. Nevertheless, even using \texttt{G2}, the gap from the mean error of raw TDOAs is always significant. The results related to \texttt{R-DEN} generally confirm what we observed for the TPR and TNR scores. Indeed, \texttt{R-DEN} is comparable to \texttt{G2} only for the cross-shaped array case, while it exhibits worse performances compared to the other methods. Finally, it can be observed that \texttt{R-DEN} achieves slightly better results when the number of outliers is very high (i.e., greater than 18). This fact is a further consequence of specifying the parameter $\kappa$ as an input of \texttt{R-DEN}. Indeed, in the rightmost part of the plots, the TPR is very high compared to that achieved by the proposed methods, i.e. \texttt{R-DEN} tends to remove more outliers than the other algorithms, thus leading to a lower residual TDOA error.

\section{Case of Study:  Acoustic Source Localization}
\label{sec:application}

In this Section we show the positive effect of TDOA outlier removal in a real-world application. More specifically, we focused on TDOA-based acoustic source localization. A localization system composed of four synchronized microphone arrays was installed in a medium-sized (i.e., $7\times 7 \times 3\, \mathrm{m}^3$) office room, with reverberation time $\mathrm{RT}_{60}\simeq0.7\,\mathrm{s}$. Within this room, we considered a localization volume of about $4 \times 2.5 \times 2\,\mathrm{m}^3$, with the microphone arrays located at four corners of the volume (two at the top and two at the bottom, at opposite sides). Each microphone array is made of four sensors, deployed on the vertices of a tetrahedron of side $40\,\mathrm{cm}$. In this scenario, a small loudspeaker emitted $30\,\mathrm{s}$ of male speech from 36 different positions homogeneously covering the localization volume. For each source position, each microphone acquired a signal at $48\,\mathrm{kHz}$. Signals were segmented into windows of $4096$ samples, from which we selected the most spectrally rich ones, to discard too harmonic-like portions of the recording. From the selected frames, we measured TDOAs within each array, by picking the maximum of the GCCs computed at all microphone pairs. Note that we did not measured inter-array TDOAs, as the microphone distance would make them not reliable \cite{Canclini2015}. Therefore, for each frame we ended up with a total of $6\times 4$ measurements. The pre-processing step (i.e., tests for $G=1$) was embedded in the computation 
of the TDOAs, as peak-picking of GCCs was restricted to the acceptance region defined in \eqref{eq:outlier_interval}.

It is worth noticing that, due to the reverberation, peak-picking of the GCCs may lead to erroneous TDOA estimation. More specifically, outlier TDOAs may be generated by two distinct phenomena (or a combination on them). On one hand, the presence of coherent reflections may introduce a bias in the position of the main peak of the GCC. On the other hand, reflections can generate secondary peaks in the GCC, which for some microphone pairs may be stronger than the expected one \cite{Omologo97}.

Table~\ref{tab:real_data_error} reports the mean TDOA error averaged over all the analyzed time windows both considering raw TDOAs and the application of outlier removal algorithms. As in this real-world case additive noise standard deviation $\sigma$ is not known, we tested outlier removal procedures using different values of $\sigma$ ranging from $0.5\,\mathrm{cm}$ to $1\,\mathrm{cm}$ (i.e., corresponding to approximately $1$ sample at $48\,\mathrm{kHz}$), which are negligible values with respect to microphone distance, thus completely fulfilling the hypothesis made in Section~\ref{sec:stat_models}. Results show that the error decreases when outliers are removed through any of the proposed procedures. When an algorithm involving \texttt{G3} exploration is used, the error is reduced by a factor greater than three. As expected from the simulative results reported in Section~\ref{sec:evaluation}, \texttt{G2} is the one with worse performances. Moreover, it is interesting to notice that a precise estimate of $\sigma$ is not mandatory, as results remain pretty stable.

\begin{table}[t]
	\centering
	\caption{Mean TDOA error [cm] obtained on real data applying outlier removal supposing different $\sigma$ values. In any case, outlier removal ensures better performances than using raw measurements.}
	\label{tab:real_data_error}
	\begin{tabular}{cccccc}
		\hline
		$\sigma$    & raw TDOAs & \texttt{G2} & \texttt{G3} & \texttt{G2+G3} & \texttt{G3+G2} \\ \hline
		$0.50\,\textrm{cm}$    & $2.02$ & $1.38$ & $0.64$ & $0.61$ & $0.64$ \\
		$0.75\,\textrm{cm}$ & $2.02$ & $1.48$ & $0.68$ & $0.66$ & $0.68$ \\ 
		$1.00\,\textrm{cm}$  & $2.02$ & $1.52$ & $0.73$ & $0.70$ & $0.73$ \\ \hline
	\end{tabular}
\end{table}

In order to investigate the positive impact of reducing the mean TDOA error, we also performed source localization for each source position. To this purpose, we iteratively minimized the Maximum-Likelihood cost function \cite{Benesty2004}, providing the room center position as the initial guess for the source location. The average Root Mean Squared Error (RMSE) is reported in Table \ref{tab:real_data_loc}. Results show source localization highly benefits from reducing the mean TDOA error. Indeed localization is more accurate using only inlier TDOAs. The best accuracy is achieved when methods involving \texttt{G3} are considered. Also in this case, localization is only slightly affected by the choice of the standard deviation $\sigma$.
 
\begin{table}[t]
	\centering
	\caption{RMSE [cm] on source localization applying outlier removal supposing different $\sigma$ values.}
	\label{tab:real_data_loc}
	\begin{tabular}{cccccc}
	\hline
	$\sigma$    & raw TDOAs & \texttt{G2} & \texttt{G3} & \texttt{G2+G3} & \texttt{G3+G2} \\ \hline
	$0.50\,\textrm{cm}$  & $38.1$ & $25.7$ & $11.6$ & $11.2$ & $11.6$  \\
	$0.75\,\textrm{cm}$ & $38.1$ & $28.0$ & $12.1$ & $11.8$ & $12.1$ \\ 
	$1.00\,\textrm{cm}$  & $38.1$ & $28.7$ & $13.4$ & $12.9$ & $13.4$ \\ \hline
\end{tabular}
\end{table}

For the sake of completeness, Table~\ref{tab:real_data_num_out} also reports the average number of rejected measurements (i.e., identified outliers) by the different tested algorithms and configurations. On average, among the $6\times 4$ available measurements, one to three of them were removed as outliers.

\begin{table}[t]
	\centering
	\caption{Average number of detected outliers over the available $6\times 4$ measurements supposing different $\sigma$ values.}
	\label{tab:real_data_num_out}
	\begin{tabular}{ccccc}
		\hline
		$\sigma$    & \texttt{G2} & \texttt{G3} & \texttt{G2+G3} & \texttt{G3+G2} \\ \hline
		$0.50\,\textrm{cm}$  & $0.89$ & $2.75$ & $3.02$ & $2.75$ \\
		$0.75\,\textrm{cm}$  & $0.75$ & $2.36$ & $2.61$ & $2.36$  \\ 
		$1.00\,\textrm{cm}$  &$0.66$ & $2.13$ & $2.36$ & $2.13$ \\ \hline
	\end{tabular}
\end{table}


\section{Conclusions}\label{sec:conclusions}

In this manuscript we proposed an algorithm for removing outliers from sets of TDOA or RD measurements. This method is developed exploiting the TDOA space framework, in which outliers can be identified as they lie too far from the feasible TDOA set. The algorithm exploits statistical testing procedures. Specifically, after a pre-processing stage, multiple testing is used to detect groups of $G=2,3$ TDOAs containing at least one outlier. Combined testing is then used to remove the outlier measurements.

The proposed approach has been first validated by means of simulations, which proved the effectiveness of the algorithm even in presence of many outliers. Moreover, as the algorithm does not  depend on any specific application, we selected the task of acoustic source localization with microphone arrays for testing the method in a real scenario. Also in this case, outlier removal showed a positive effect, significantly improving the localization accuracy.


\ifCLASSOPTIONcaptionsoff
  \newpage
\fi

\balance

\bibliographystyle{IEEEtran}
\bibliography{biblio}{}

\end{document}